\documentclass[aps,prd,reprint,showpacs,floatfix,a4paper,superscriptaddress]{revtex4-1}
\usepackage{graphicx}
\usepackage{amsmath}
\usepackage{amssymb}
\usepackage{amsfonts}
\usepackage{color}
\usepackage{relsize}
\usepackage{mathrsfs}
\usepackage{txfonts}
\usepackage[T1]{fontenc}
\usepackage{microtype}
\usepackage[pdfpagemode=UseNone,colorlinks=true,linkcolor=blue,citecolor=blue]{hyperref}

\renewcommand\Re{\operatorname{Re}}
\renewcommand\Im{\operatorname{Im}}

\DeclareMathOperator{\sign}{sign}

\newcommand{\ha}{a}
\newcommand{\omm}{\omega_{\mathrm{m}}}
\newcommand{\hp}{p}
\newcommand{\hq}{q}
\newcommand{\Gz}{G_{0}}
\newcommand{\ii}{\mathrm{i}}
\newcommand{\EE}{\mathscr{E}}

\newcommand{\ep}{\epsilon}
\newcommand{\gam}{\gamma_{\mathrm{m}}}

\newcommand{\ka}{\kappa}
\newcommand{\De}{\varDelta}
\newcommand{\Dez}{\De_{0}}
\newcommand{\phd}{\dot{\phi}}

\newcommand{\al}{\alpha}
\newcommand{\qs}{q_{\mathrm{s}}}
\newcommand{\dhq}{\delta\hq}
\newcommand{\ps}{p_{\mathrm{s}}}
\newcommand{\dhp}{\delta\hp}
\newcommand{\als}{\al_{\mathrm{s}}}
\newcommand{\alss}{\al_{\mathrm{s}}^{*}}
\newcommand{\dha}{\delta\ha}
\newcommand{\ddhq}{\delta\dot{q}}
\newcommand{\ddhp}{\delta\dot{p}}
\newcommand{\ddha}{\delta\dot{a}}
\newcommand{\dhad}{\delta a^{\dag}}

\newcommand{\NML}[1]{\nonumber\\&&#1}
\newcommand{\om}{\omega}

\newcommand{\rmi}{\mathrm{i}}
\newcommand{\rme}{\mathrm{e}}
\newcommand{\rmd}{\mathrm{d}}
\newcommand{\rmm}{\mathrm{m}}
\newcommand{\rmc}{\mathrm{c}}
\newcommand{\rmin}{\mathrm{in}}
\newcommand{\rmout}{\mathrm{out}}
\newcommand{\mce}{\mathscr{E}}
\newcommand{\myS}[1]{\operatorname{S}_{#1}}
\newcommand{\mySbar}[1]{\bar{\operatorname{S}}_{#1}}

\begin{document}

\title{Frequency noise cancellation in optomechanical systems for ponderomotive squeezing}

\author{A.~Pontin}
\affiliation{Istituto Nazionale di Fisica Nucleare (INFN), Gruppo Collegato di Trento, I-38123 Povo (TN), Italy}
\affiliation{Dipartimento di Fisica, Universit\`a di Trento, I-38123 Povo (TN), Italy}
\author{C.~Biancofiore}
\affiliation{School of Science and Technology, Physics Division, University of Camerino, via Madonna delle Carceri, 9, I-62032 Camerino (MC), Italy, and INFN, Sezione di Perugia, Italy}
\author{E.~Serra}
\affiliation{Istituto Nazionale di Fisica Nucleare (INFN), Gruppo Collegato di Trento, I-38123 Povo (TN), Italy}
\affiliation{Interdisciplinary Laboratory for Computational Science (LISC), FBK-University of Trento, I-38123 Povo (TN), Italy}
\affiliation{Dept. of Microelectronics and Computer, Engineering/ECTM/DIMES, Feldmanweg, 17-2628 CT Delft, PO Box 5053, 2600 GB Delft, The Netherlands }

\author{A.~Borrielli}
\affiliation{Istituto Nazionale di Fisica Nucleare (INFN), Gruppo Collegato di Trento, I-38123 Povo (TN), Italy}
\affiliation{Institute of Materials for Electronics and Magnetism, Nanoscience-Trento-FBK Division, I-38123 Povo (TN), Italy}
\author{F.~S.~Cataliotti}
\affiliation{Dipartimento di Fisica e Astronomia, Universit\`a di Firenze, Via Sansone 1, I-50019 Sesto Fiorentino (FI), Italy}
\affiliation{European Laboratory for Non-Linear Spectroscopy (LENS), Via Carrara 1, I-50019 Sesto Fiorentino (FI), Italy}
\affiliation{INFN, Sezione di Firenze, Via Sansone 1, I-50019 Sesto Fiorentino (FI), Italy}
\author{F.~Marino}
\affiliation{European Laboratory for Non-Linear Spectroscopy (LENS), Via Carrara 1, I-50019 Sesto Fiorentino (FI), Italy}
\affiliation{INFN, Sezione di Firenze, Via Sansone 1, I-50019 Sesto Fiorentino (FI), Italy}
\affiliation{CNR-Istituto Nazionale di Ottica, Via Carrara 1, I-50019 Sesto Fiorentino (FI), Italy}

\author{G.~A.~Prodi}
\affiliation{Istituto Nazionale di Fisica Nucleare (INFN), Gruppo Collegato di Trento, I-38123 Povo (TN), Italy}
\affiliation{Dipartimento di Fisica, Universit\`a di Trento, I-38123 Povo (TN), Italy}
\author{M.~Bonaldi}
\affiliation{Istituto Nazionale di Fisica Nucleare (INFN), Gruppo Collegato di Trento, I-38123 Povo (TN), Italy}
\affiliation{Institute of Materials for Electronics and Magnetism, Nanoscience-Trento-FBK Division, I-38123 Povo (TN), Italy}
\author{F.~Marin}\email[Corresponding author:] {marin@fi.infn.it}
\affiliation{Dipartimento di Fisica e Astronomia, Universit\`a di Firenze, Via Sansone 1, I-50019 Sesto Fiorentino (FI), Italy}
\affiliation{European Laboratory for Non-Linear Spectroscopy (LENS), Via Carrara 1, I-50019 Sesto Fiorentino (FI), Italy}
\affiliation{INFN, Sezione di Firenze, Via Sansone 1, I-50019 Sesto Fiorentino (FI), Italy}
\author{D.~Vitali}\email[Corresponding author:] {david.vitali@unicam.it}
\affiliation{School of Science and Technology, Physics Division, University of Camerino, via Madonna delle Carceri, 9, I-62032 Camerino (MC), Italy, and INFN, Sezione di Perugia, Italy}

\date{\today}

\begin{abstract}
Ponderomotive squeezing of the output light of an optical cavity has been recently observed in the MHz range in two different cavity optomechanical devices. Quadrature squeezing becomes particularly useful at lower spectral frequencies, for example in gravitational wave interferometers, despite being more sensitive to excess phase and frequency noise. Here we show a phase/frequency noise cancellation mechanism due to destructive interference which can facilitate the production of ponderomotive squeezing in the kHz range and we demonstrate it experimentally in an optomechanical system formed by a Fabry-P\'{e}rot cavity with a micro-mechanical mirror.
\end{abstract}


\maketitle

\section{Introduction}

Optical squeezing is an important tool for improving quantum limited displacement sensing~\cite{Braginsky1995}, as first pointed out in
proposals to increase the displacement sensitivity of large scale gravitational wave observatories~\cite{Caves1981,Kimble2001}. Squeezed light was first produced using atomic sodium as a nonlinear medium \cite{Slusher1985}, and then in experiments employing optical fibers \cite{Shelby1986} and nonlinear crystals \cite{Wu1986}. Substantial squeezing has been achieved in modern experiments (up to 12.7 dB~\cite{Eberle2010}), and enhanced sensitivity using squeezed light has been realized in gravitational wave detectors \cite{LIGO2011} and in biological measurements \cite{Taylor2013}.

Searches for ever-better squeezing materials led to suggest the possibility to generate ponderomotive squeezing~\cite{Mancini1994,Fabre1994}, i.e., quadrature-squeezed light at the output of a cavity caused by the radiation pressure interaction of the cavity mode with a vibrating resonator. The mechanical element is shifted proportionally to the intracavity intensity, and consequently the optical path inside the cavity depends upon such intensity. Therefore the optomechanical system behaves similarly to a cavity filled with a nonlinear Kerr medium, however, with two important differences: (i) the effective nonlinearity is delayed by a time depending upon the dynamics of the mechanical element; (ii) the optomechanical interaction transmits mechanical thermal noise to the cavity field, causing fluctuations of its frequency. When the mechanical oscillator is fast enough, i.e., we look at frequencies much lower than the mechanical resonance, the mechanical response is instantaneous, and the optomechanical interaction becomes indistinguishable from a Kerr interaction.
It is known that when a cavity containing a Kerr medium is driven by an intense laser, one gets appreciable squeezing
in the spectrum of quadrature fluctuations at the cavity output \cite{Walls1995}. The above analogy therefore suggests that a strongly driven optomechanical cavity will also be able to produce quadrature squeezing at its output, provided that optomechanical coupling predominates over the detrimental effect of thermal noise. The first analysis of Refs.~\cite{Mancini1994,Fabre1994} was then later extended to the case of many vibrational modes in \cite{Pinard1999}. The problem was then reconsidered in a Michelson interferometer setup in \cite{Corbitt2006}, and an experimental study of the possible signatures of ponderomotive squeezing in a Fabry-P\'{e}rot cavity with a movable end-mirror was then carried out in \cite{Marino2010}.
More recently ponderomotively squeezed light at the few percent level has been demonstrated using a mechanical mode of an ultracold atomic gas inside an optical cavity \cite{Brooks2012}, and more recently using a silicon micromechanical resonator~\cite{Safavi-Naeini2013}, and a thin semi-transparent membrane within a Fabry-P\'{e}rot cavity~\cite{Purdy2013}. These latter experiments achieved squeezing around the mechanical resonance in the MHz range; however quadrature squeezing is particularly useful for improving sensitivity at lower frequencies, in the audio-band, for example for improving the sensitivity of gravitational wave interferometers~\cite{Kimble2001}. At lower frequencies however, various sources of technical noise, such as thermal noise, phase/frequency noise associated with the input field and/or the slow cavity fluctuations, have detrimental effects on squeezing, making low-frequency ponderomotive squeezing much more difficult to achieve.

Here we consider the general problem of ponderomotive squeezing in the presence of thermal, intensity and frequency noise. We show that frequency noise can be almost completely cancelled around the \emph{bare} mechanical resonance frequency due to the destructive interference between the frequency noise directly affecting the cavity and the same frequency noise transduced by the mechanical resonator. We demonstrate such an effect experimentally in a frequency band around 100 kHz, using an optomechanical setup formed by a Fabry-P\'{e}rot cavity with a micromechanical mirror~\cite{Serra2012}. Finally we also show that such a noise cancellation could facilitate the generation and detection of ponderomotive squeezing in the audio-band.

The paper is as follows. In Sec.~II we provide a general description of an optomechanical system subject both to quantum and technical noise in terms of quantum Langevin equations. In Sec.~III we calculate the output quadrature noise spectra and illustrate the frequency noise cancellation mechanism, also by comparing it with different destructive interference phenomena such as optomechanical induced transparency (OMIT)~\cite{Weis2010,Safavi-Naeini2011,Teufel2011,Karuza2013}, and backaction amplification~\cite{Buonanno2001,Arcizet2006,Verlot2010}. In Sec.~IV we describe the experimental cavity optomechanical apparatus, and in Sec.~V we illustrate the experimental results. Finally in Sec.~VI we show that this noise cancellation may facilitate the generation and detection of ponderomotive squeezing in the kHz range and using microgram mechanical resonators.

\section{Model}

We consider a generic cavity optomechanical system in which a mechanical resonator with frequency $\omega_{\rmm}$ is subject to a force proportional to the photon number of an optical cavity mode with frequency $\omega_{\rmc}$, which is driven by an intense laser.
The corresponding Hamiltonian can be written as \cite{Kippenberg2007,Genes2009,Law1995,Giovannetti2001}
\begin{eqnarray}
H&=&\hbar\omega_{\rmc}a^{\dagger}a+\frac{1}{2}\hbar\omega_{\rmm}(p^{2}+q^{2})
-\hbar G_{0}a^{\dagger}a q \nonumber \\
&&+\rmi\hbar \mce_0 \bigl(a^{\dagger}\rme^{-\rmi\omega_{0}t}-a\,\mathrm{e}^{\rmi\omega_{0}t}\bigr).
\label{hamiltonian}
\end{eqnarray}
The first term describes the energy of the cavity mode, with annihilation operator $a$ ($[a,a^{\dag}]=1$), while the second term gives the energy of the mechanical resonator, described by dimensionless position and momentum operators $q$ and $p$, satisfying the commutation relation $[q,p]=\rmi$.
The third term is the optomechanical interaction, with single photon optomechanical coupling strength
\begin{equation}\label{eq:coupling}
    G_0=-\biggl(\frac{\rmd \omega_{\rmc}}{\rmd x}\biggr)\sqrt{\frac{\hbar}{m \omega_{\rmm}}},
\end{equation}
where $(\rmd \omega_{\rmc}/\rmd x)$ is the change in cavity frequency per displacement and $m$ is the effective mass of the mechanical mode.
The last term describes the cavity driving by a laser with frequency $\omega_0$ and $\mce_0=\sqrt{2 \kappa_1 \mathscr{P}/\hbar \omega_0}$, where $\mathscr{P}$ is the input laser power and $\kappa_1$ is the cavity loss rate through its input port. We have taken the amplitude $\mce_0$ real, which means that we use the driving laser as phase reference for the optical field.

For a full description of the system dynamics it is necessary to include the fluctuation-dissipation processes affecting both the optical and the mechanical mode. They can be taken into account in a fully consistent way \cite{Giovannetti2001}, and in the frame rotating at the laser frequency $\omega_0$ one gets
\begin{subequations}
\label{nonlinear0}
\begin{eqnarray}
\dot{q}&=&\omega_{\rmm} p, \\
\dot{p}&=&-\omega_{\rmm} q - \gamma_{\rmm} p + G_0 a^{\dag}a + \xi, \\
\dot{a}&=&-\kappa a-\rmi\bigl[\omega_{\rmc}-\omega_0-G_0 q\bigr]a \nonumber \\
&&+\mce_0+\sqrt{2\kappa_1} a_1^{\rmin}+\sqrt{2\kappa_2} a_2^{\rmin}.
\end{eqnarray}
\end{subequations}
$a_1^{\rmin}$ is the vacuum input noise entering the input port of the optomechanical cavity, $a_2^{\rmin}$ is the vacuum input noise describing all the other decay channels (optical losses and transmission through the back mirror), with decay rate $\kappa_2$, and $\kappa=\kappa_1+\kappa_2$ is the total cavity decay rate. The two input noises are uncorrelated and possess the following correlation functions~\cite{Gardiner2000}
\begin{subequations}
\label{input}
\begin{eqnarray}
&& \bigl\langle a_j^{\rmin}(t)a_j^{\rmin}(t')\bigr\rangle = \bigl\langle a_j^{\rmin, \dag}(t)a_j^{\rmin, \dag}(t')\bigr\rangle = \bigl\langle a_j^{\rmin,\dag}(t)a_j^{\rmin}(t')\bigr\rangle= 0,\quad\\
&& \bigl\langle a_j^{\rmin}(t)a_j^{\rmin,\dag}(t')\bigr\rangle = \delta (t-t'), \, j=1,2.
\end{eqnarray}
\end{subequations}
The mechanical mode is affected by a viscous force with damping rate $\gamma_{\rmm}$ and by a Brownian stochastic force with zero mean value $\xi(t)$, obeying the correlation function at temperature $T$ \cite{Landau1958,Gardiner2000}
\begin{equation}
\bigl \langle \xi(t) \xi(t')\bigr \rangle = \frac{\gamma_{\rmm}}{\omega_{\rmm}}%
\int \frac{\rmd \omega}{2\pi} \rme^{-\rmi \omega(t-t')} \omega \biggl[\coth\biggl(\frac{\hbar \omega}{2k_{\mathrm{B}}T}\biggr)+1\biggr].
\label{brownian}
\end{equation}
The Brownian noise $\xi(t)$ is a Gaussian quantum stochastic process and its non-Markovian nature (neither its correlation function nor its commutator are proportional to a Dirac delta function) guarantees that the Quantum Langevin Equations (QLE) of Eqs.~(\ref{nonlinear}) preserve the correct commutation relations between operators during the time evolution \cite{Giovannetti2001}.

The above mentioned thermal and vacuum noises are unavoidable fundamental noises. However in a realistic scenario two additional technical noises (uncorrelated between them and with all the other noises) can play a relevant role: (i) amplitude noise, which is taken into account considering that the coherent amplitude $\mce_0$ at the input of the cavity mode is fluctuating, $\mce_0 \to \mce_0 + \sqrt{2\kappa_1}\epsilon(t)$, where $\epsilon(t)$ is a real, zero-mean Gaussian stochastic variable; (ii) phase/frequency noise, which is caused both by the laser frequency fluctuations, and by the fluctuations of the cavity length (and therefore of its resonance frequency) which are not due to the considered mode of the mechanical resonator. This noise manifests itself as a fluctuating detuning $\omega_{\rmc}-\omega_0 \to \varDelta_0 -\dot{\phi}(t)$, where $\dot{\phi}(t)$ is a zero-mean frequency noise. As a
 consequence Eqs.~(\ref{nonlinear0}) become
\begin{subequations}
\label{nonlinear}
\begin{eqnarray}
\dot{q}&=&\omega_{\rmm} p, \\
\dot{p}&=&-\omega_{\rmm} q - \gamma_{\rmm} p + G_0 a^{\dag}a + \xi, \\
\dot{a}&=&-\kappa a -\rmi\bigl(\varDelta_0-\dot{\phi}-G_0 q\bigr)a   +\mce_0 \nonumber \\
&& +\sqrt{2\kappa_1}\left( a_1^{\rmin}+\epsilon\right)+\sqrt{2\kappa_2} a_2^{\rmin}.
\end{eqnarray}
\end{subequations}
Therefore in the frame rotating at the fluctuating frequency, amplitude noise acts as additive noise on the cavity modes, while frequency noise is a multiplicative noise, affecting the cavity field in the same manner of the fluctuations of the resonator position $q$.

We want to generate and manipulate optical quantum fluctuations and therefore we consider the motion of the system around a steady state characterized by the intracavity electromagnetic field in an approximate coherent state of amplitude $\al_{\mathrm{s}}$, and the micro-oscillator at a new position $q_{\mathrm{s}}$, by writing:
\begin{eqnarray}
 \hq&=&\qs+\dhq,\label{eq:ss+fluct_q}\\
 \hp&=&\ps+\dhp,\label{eq:ss+fluct_p}\\
 \ha&=&\als+\dha.\label{eq:ss+fluct_a}
\end{eqnarray}
Substituting Eqs.~(\ref{eq:ss+fluct_q})-(\ref{eq:ss+fluct_a}) into Eq.~(\ref{nonlinear}), and retaining only the 0-th order contributions one gets:
\begin{eqnarray}\label{eq:ss_0th}
 \qs&=&\frac{\Gz}{\omm}\lvert\als\rvert^{2},\\
 \ps&=&0,\\
 \als&=&\frac{\EE_{0}}{\ka+\ii\De}, \label{eq:ss_1th}
\end{eqnarray}
where $\De=\Dez-\Gz\qs=\Dez-\Gz^{2}\lvert\als\rvert^{2}/\omm$.

The exact and non-linear QLE for the fluctuation operators are given by:
\begin{subequations}
\label{exact}
\begin{eqnarray}
\ddhq&=&\omm\dhp,\\
 \ddhp&=&-\omm\dhq-\gam\dhp+\Gz\bigl(\als\dhad+\alss\dha\bigr)+\xi \NML{+}\Gz\dhad\dha,\\
 \ddha&=&-\bigl(\ka+\ii\De\bigr)\dha+\ii\Gz\als\dhq+\ii\bigl(\Gz\dhq+\phd\bigr)\dha+\ii\phd\als \nonumber \\
 &&+\sqrt{2\kappa_1} \left( a_1^{\rmin}+\epsilon\right)+\sqrt{2\kappa_2} a_2^{\rmin}.
\end{eqnarray}
\end{subequations}
The nonlinear terms are $\Gz\dhad\dha$, $\ii\Gz\dhq \dha$, and $\ii\Gz\phd\dha$. The first two terms have negligible effect when $|\als | \gg 1$, which is usually satisfied, and therefore they can be safely neglected. The last term is a multiplicative noise term and it is not obvious if and when it can be neglected since its evaluation requires the knowledge (or realistic hypotheses) of the frequency and displacement noise spectrum on a wide frequency range. Its treatment is outside the purpose of the present work and we shall neglect this last term in the following. Therefore Eqs.~(\ref{exact}) become
\begin{subequations}
\label{linear}
\begin{eqnarray}
\ddhq&=&\omm\dhp,\\
 \ddhp&=&-\omm\dhq-\gam\dhp+\Gz\bigl(\als\dhad+\alss\dha\bigr)+\xi,\\
 \ddha&=&-\bigl(\ka+\ii\De\bigr)\dha+\ii\Gz\als\dhq+\sqrt{2\kappa_1} \tilde{a}_1^{\rmin}+\zeta,
\end{eqnarray}
\end{subequations}
where we have introduced the two noise terms
\begin{eqnarray}\label{eq:zeta}
    \zeta & = & \ii\als\phd+\sqrt{2\kappa_2} a_2^{\rmin},\\
    \tilde{a}_1^{\rmin}& = & a_1^{\rmin}+\ep. \label{eq:atilde}
\end{eqnarray}

\section{Output quadrature spectra}

Ponderomotive squeezing is detected in the noise spectrum of appropriate quadratures of the optical cavity output. This noise spectrum can be calculated from the Fourier transform of the field at the input/output port,
\begin{equation}\label{eq:output}
    a_1^{\rmout}=\sqrt{2\kappa_1}\dha-\tilde{a}_1^{\rmin}.
\end{equation}
Taking the Fourier transform of Eqs. (\ref{linear}), solving for $\dha(\omega)$ and using Eq.~(\ref{eq:output}), one gets
\begin{eqnarray}\label{eq:aout}
  a_1^{\rmout}(\omega) &=& \nu_1(\omega)\tilde{a}_1^{\rmin}(\omega) + \nu_2(\omega)\tilde{a}_1^{\rmin,\dag}(\omega)\nonumber \\
   &&+ \nu_3(\omega)\zeta(\omega) + \nu_4(\omega)\zeta^{\dag}(\omega) +\nu_{\mathrm{T}}(\omega)\xi(\omega),
\end{eqnarray}
where
\begin{subequations}
\label{eq:nu}
\begin{eqnarray}\label{eq:nu1}
  \nu_1(\omega) &=& \frac{(1-2\eta)\kappa-\ii\bigl(\varDelta-\omega\bigr)}{\kappa+\ii\bigl(\varDelta-\omega\bigr)}+\frac{\ii |G|^2 \kappa_1 \chi_\mathrm{eff}(\omega)}{\bigl[\kappa+\ii\bigl(\varDelta-\omega\bigr)\bigr]^2}, \\
  \label{eq:nu2}
  \nu_2(\omega) &=& \frac{\ii G^2 \kappa_1 \chi_\mathrm{eff}(\omega)}{\bigl[\kappa+\ii\bigl(\varDelta-\omega\bigr)\bigr]\bigl[\kappa-\ii\bigl(\varDelta+\omega\bigr)\bigr]}, \\
  \label{eq:nu3}
  \nu_3(\omega) &=& \frac{\sqrt{2\kappa_1}}{\kappa+\ii\bigl(\varDelta-\omega\bigr)}\Biggl\{1+\frac{\ii |G|^2 \chi_\mathrm{eff}(\omega)}{2\bigl[\kappa+\ii\bigl(\varDelta-\omega\bigr)\bigr]}\Biggr\} \nonumber \\ &=&\frac{\nu_1(\om)+1}{\sqrt{2\kappa_1}}, \\
  \label{eq:nu4}
  \nu_4(\omega) &=& \frac{\nu_2(\om)}{\sqrt{2\kappa_1}}, \\
   \label{eq:nu5}
  \nu_{\mathrm{T}}(\omega) &=& \frac{\ii G \sqrt{\kappa_1} \chi_\mathrm{eff}(\omega)}{\kappa+\ii\bigl(\varDelta-\omega\bigr)},
  \end{eqnarray}
  \end{subequations}
where $\eta = \kappa_2/\kappa$, ($0\leq \eta \leq 1$) is a parameter quantifying optical losses, $G=G_0 \sqrt{2} \als$ is the effective coupling, and
\begin{equation}
\chi _\mathrm{ eff}(\omega )=\omega _\mathrm{m}\Biggl[\omega_\mathrm{m}^{2}-\omega^{2}-\mathrm{i}\omega \gamma _\mathrm{m}-\frac{|G|^2\varDelta\omega _\mathrm{m}}{\bigl(\kappa -\mathrm{i}\omega \bigr)^{2}+\varDelta^{2}}\Biggr]^{-1} ,\label{chieffD}
\end{equation}%
is the effective mechanical susceptibility modified by the optomechanical coupling.
The first two functions $\nu_1(\omega)$ and $\nu_2(\omega)$ are responsible for the generation of ponderomotive squeezing, while the last three functions determine the response of the optomechanical system to frequency noise, thermal noise, and optical losses.

\subsection{Frequency noise cancellation}
\label{IIIA}

Eq.~(\ref{eq:aout}) already enables to predict that an interesting frequency noise cancellation takes place. In fact, using Eq.~(\ref{eq:zeta}), one has for the contribution to the output signal proportional to $\dot{\phi}$
\begin{equation}\label{eq:cancel}
     a_1^{\rmout}(\omega) = \mathrm{i} \bigl[\als\nu_3(\omega)-\als^*\nu_4(\omega)\bigr]\dot{\phi}(\omega) \propto \frac{\chi _\mathrm{ eff}(\omega )}{\chi _{0}(\omega )} \dot{\phi}(\omega),
\end{equation}
where
\begin{equation}
\chi _{0}(\omega )=\omega _\mathrm{m}\Bigl(\omega_\mathrm{m}^{2}-\omega^{2}-\mathrm{i}\omega \gamma _\mathrm{m}\Bigr)^{-1} ,\label{chi0}
\end{equation}
is the bare susceptibility of the mechanical resonator. In the usual case of a large mechanical quality factor, at the unperturbed mechanical resonance $\omega = \omega_{\mathrm{m}}$, $\chi _{0}(\omega )$ diverges and \emph{the output signal is unaffected by frequency noise}. Therefore we expect a narrow bandwidth around $\omega = \omega_{\mathrm{m}}$ where noise is strongly suppressed. The mechanism at the basis of this cancellation is the interference between the frequency noise directly affecting the cavity and the same frequency noise transduced by the mechanical resonator, which is perfectly destructive when $\omega = \omega_{\mathrm{m}}$. This cancellation takes place at any cavity detuning except at resonance $\varDelta =0$, when $\chi_\mathrm{eff}(\omega )=\chi _{0}(\omega )$. In this case in fact, the mechanical resonator is sensitive only to intensity noise and cannot transduce phase/frequency noise.

This frequency noise cancellation is related to the backaction amplification of a signal discussed in Refs.~\cite{Buonanno2001,Arcizet2006} and demonstrated in Ref.~\cite{Verlot2010}. In fact, both effects are described by the ``amplification'' ratio $ \chi _\mathrm{ eff}(\omega )/\chi _{0}(\omega )$, because the system responds in the same way to cavity length variations due either to an external signal or to frequency noise modulations. However, here we exploit this interference phenomenon for a different purpose, i.e., for reducing phase/frequency noise in the optical output rather than for amplifying an external signal. We shall see in Sec.~VI that such a noise cancellation is essential for the possibility of generating and detecting ponderomotive squeezing at hundreds of kHz.

This frequency noise cancellation is instead different from the optomechanical induced transparency (OMIT) of an optical probe beam, which is caused by the destructive interference between the probe itself and the blue motional sideband of the driving pump beam~\cite{Weis2010,Safavi-Naeini2011,Teufel2011,Karuza2013}. Indeed, in OMIT the probe transmission is described by the coefficient $\nu_3(\omega)$ of Eq.~(\ref{eq:nu3}), which can be rewritten as
\begin{equation}
\nu_3(\omega) = \frac{\sqrt{2\kappa_1}}{\kappa+\ii\bigl(\varDelta-\omega\bigr)}\chi_\mathrm{eff}(\omega)\Biggl\{\chi_\mathrm{0}(\omega)^{-1}+\frac{\ii |G|^2}{2\bigl[\kappa-\ii\bigl(\varDelta+\omega\bigr)\bigr]}\Biggr\},
\end{equation}
and therefore is not proportional to $ \chi _\mathrm{ eff}(\omega )/\chi _{0}(\omega )$, and never reduces to it. In fact, when the last term on the right hand side, proportional to $|G|^2$ is negligible, $ \chi _\mathrm{ eff}(\omega )/\chi _{0}(\omega )$ simultaneously becomes essentially equal to one, and therefore also frequency noise cancellation disappears.
The difference between OMIT and the present frequency noise cancellation is manifested also by the fact that in OMIT destructive interference (that is $\nu_3(\omega)\simeq 0$) occurs at $\omega \simeq \omega_{\rm m}^{\rm eff}$, the effective mechanical frequency shifted by the optical spring effect. On the contrary, frequency noise cancellation occurs when $ \chi _\mathrm{ eff}(\omega )/\chi _{0}(\omega \simeq 0$, which occurs at the \emph{bare} mechanical frequency $\omega_{\rm m}$, while at $\omega_{\rm m}^{\rm eff}$ such a ratio is maximum.

\begin{figure}[h]
 \centering
\includegraphics[width=0.375\textwidth]{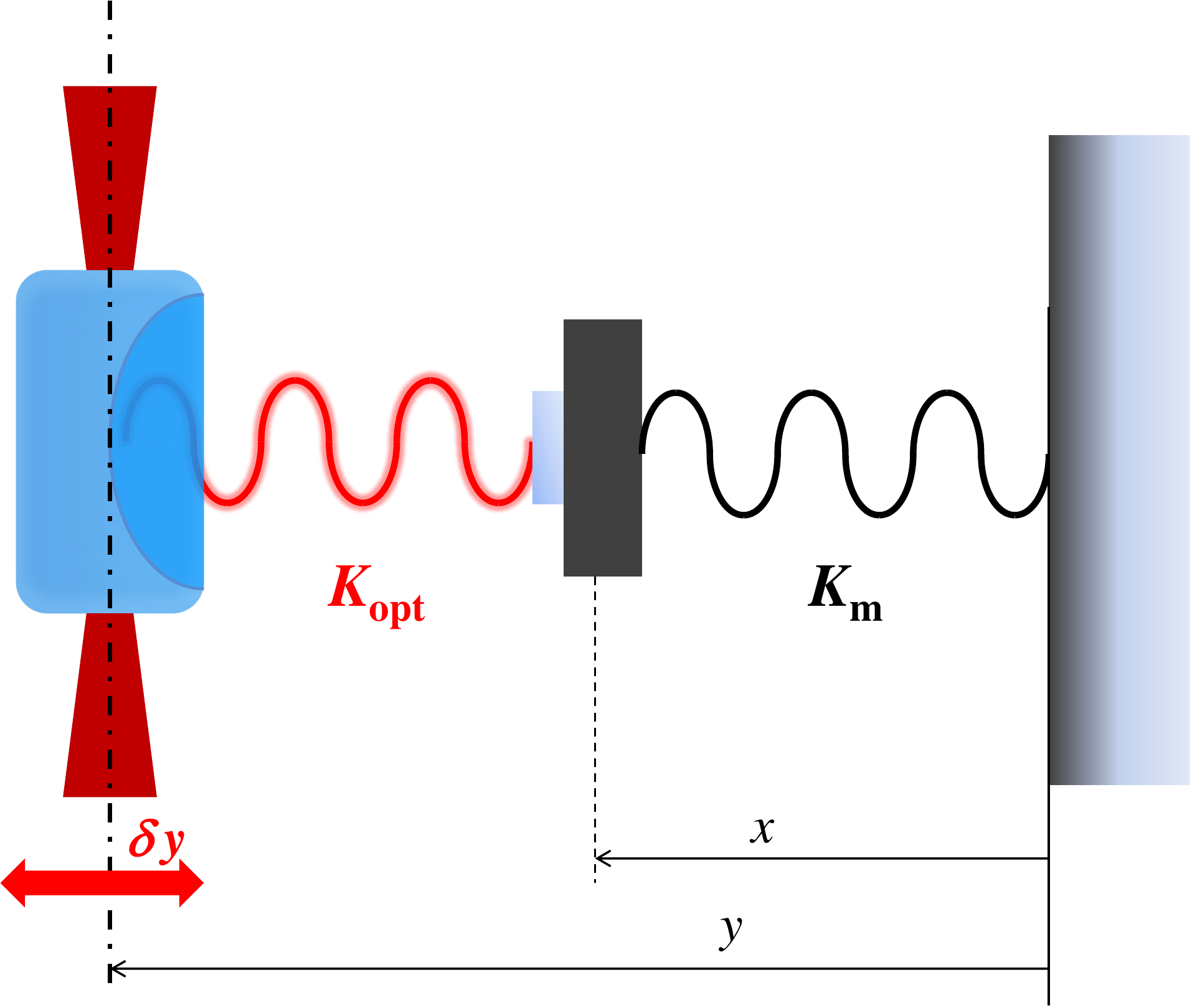}
\caption{(Color online) Scheme of the simplified model explaining the effect of noise cancellation. The oscillating mass (micro-mirror) is linked by a mechanical spring ($K_{\mathrm{m}}$) to a fixed frame, and by the optical spring ($K_{\mathrm{opt}}$) to the input mirror, modeled as a rigid fluctuating bound.}
 \label{modellino}
\end{figure}

The physical origin of the phenomenon of noise cancellation can be understood also with a simple model. As we have seen, a single classic variable (in our notation, $\dot{\phi}$) describes the fluctuations in the cavity detuning, and it can be used to take into account both the laser frequency noise and the cavity length fluctuations (excluding the oscillator modes with low effective mass, for which it is necessary to include in the description their response to radiation pressure). As a consequence, we can consider in this simple model all such noise sources as contributions to effective position fluctuations of the input cavity mirror. Around the mechanical resonance (when $\omega \simeq \omega_\mathrm{m}$) the effective susceptibility can be expressed by defining an \emph{optical spring} with strength
\begin{equation}
K_{\mathrm{opt}} = m \,\omega_{\mathrm{opt}}^2 = \frac{m \,|G|^2\varDelta\omega _{\mathrm{m}}\,\bigl(\kappa^2+\varDelta^2-\omega_{\mathrm{m}}^2\bigr)}{\bigl(\kappa^{2}+\varDelta^{2}-\omega_{\mathrm{m}}^2\bigr)^2+4\kappa^2\omega_{\mathrm{m}}^2}
\end{equation}
and damping
\begin{equation}
\gamma_\mathrm{opt} = \frac{2 \kappa |G|^2\varDelta\omega _{\mathrm{m}}}{\bigl(\kappa^{2}+\varDelta^{2}-\omega_{\mathrm{m}}^2\bigr)^2+4\kappa^2\omega_{\mathrm{m}}^2} \, ,
\end{equation}
that allows to write
\begin{equation}
\chi _\mathrm{ eff}(\omega ) \simeq \omega _\mathrm{m}\Bigl[\bigl(\omega_\mathrm{m}^{2}-\omega_\mathrm{opt}^2\bigr)-\omega^{2}-\mathrm{i}\omega \bigl(\gamma _\mathrm{m}+\gamma _\mathrm{opt}\bigr)\Bigr]^{-1} .
\end{equation}
The overall system (cavity with oscillating micro-mirror feeling radiation pressure) can be sketched (see Fig.~\ref{modellino}) as a first mirror with a fluctuating position $y(t)$ (we remark that we are not considering a noise force on the first mirror, but a position noise), and a second mirror (the micro-oscillator) linked by the optical spring of strength $K_{\mathrm{opt}}$ to the first mirror, and by the mechanical spring $K_{\mathrm{m}}$ to a fix rigid frame. The second mirror has mass $m$ and a fluctuating position $x(t)$ around equilibrium. Neglecting the damping, the equation of motion for $x$ is
\begin{equation}
m \ddot{x} + K_{\mathrm{m}} x -K_{\mathrm{opt}} (y-x) = 0
\end{equation}
giving the solution, for the Fourier-transformed variables $\tilde{x}(\omega)$ and $\tilde{y}(\omega)$,
\begin{equation}
 \tilde{x}=\frac{K_{opt}}{K_{m}+K_{opt}-m \omega^{2}}\tilde{y}=\frac{\omega_{opt}^{2}}{\omega_{m}^{2}+\omega_{opt}^{2}-\omega^{2}}\tilde{y}
\end{equation}
and for the distance $(y-x)$ (corresponding to the cavity detuning in the real system)
\begin{equation}
\tilde{y}-\tilde{x}=\frac{\omega_{m}^{2}-\omega^{2}} {\omega_{m}^{2}+\omega_{opt}^{2}-\omega^{2}}\tilde{y}=\frac{\chi_{eff}}{\chi_{0}} \tilde{y}.
\end{equation}
We have therefore a cancelation of the effect of the mirror position noise on the cavity length when $\chi_{0}\gg \chi_{eff}$, i.e., around the bare oscillator resonance. The cancelation on the cavity detuning is effective also on the intracavity and output fields.

In the next Section we will experimentally demonstrate such a cancellation of frequency noise in a cavity optomechanical setup.

\subsection{Noise spectra}

In general, we detect the noise spectrum of the quadrature at phase $\varphi$ ($0\leq \varphi < \pi$), $d_{\varphi}= a^{\mathrm{out}}_{1}e^{-\ii \varphi}+a^{\mathrm{out},\dag}_{1}e^{\ii \varphi}$ which, due to stationarity, is defined as
\begin{equation} \label{eq:spefi}
 2\pi \myS{d}^{\varphi}(\om)\delta(\om+\om')= \frac{\bigl\langle d_{\varphi}(\om) d_{\varphi}(\om')\bigr\rangle+\bigl\langle d_{\varphi}(\om') d_{\varphi}(\om)\bigr\rangle}{2}.
\end{equation}
The output light is squeezed at phase $\varphi$ when the noise spectrum is below the shot-noise limit, i.e., when $\myS{d}^{\varphi}(\om)<1$ in our definitions. The quadrature noise spectrum $\myS{d}^{\varphi}(\om)$ can be written in terms of the noise spectra of the amplitude ($X^\mathrm{ out}\equiv d_0$) and
phase ($Y^\mathrm{ out}\equiv d_{\pi/2}$) quadratures, $\myS{X}(\om)$ and $\myS{Y}(\om)$ respectively, and their symmetrized correlation spectrum $\myS{X,Y}(\om)$, as
\begin{eqnarray}\label{eq:Sdphi}
 \myS{d}^{\varphi}(\om)&=&\frac{\myS{X}(\om)+\myS{Y}(\om)}{2}+\frac{\myS{X}(\om)-\myS{Y}(\om)}{2}\cos(2\varphi) \NML{+}\myS{X,Y}(\om)\sin(2\varphi).
\end{eqnarray}
We recall that the Heisenberg's uncertainty relation implies for these noise spectra
\begin{equation}\label{eq:Heisenberg}
 \myS{X}(\om)\myS{Y}(\om)-[\myS{X,Y}(\om)]^{2}\ge 1.
\end{equation}
The optimal squeezing spectrum $\myS{d}^{\mathrm{opt}}(\om)$ and the corresponding frequency-dependent optimal phase $\varphi^{\mathrm{opt}}_{d}(\om)$ of the output mode can be calculated by minimizing $\myS{d}^{\varphi}(\om)$ with respect to the phase $\varphi$, obtaining:
\begin{eqnarray}
 2 \myS{d}^{\mathrm{opt}}(\om)&=&\myS{X}(\om)+\myS{Y}(\om)-\Bigl\{\bigl[\myS{X}(\om)-\myS{Y}(\om)\bigr]^{2} \NML{+}4\bigl[\myS{X,Y}(\om)\bigr]^{2}\Bigr\}^{1/2},\\
 2\,\varphi^{\mathrm{opt}}_{d}(\om)&=&\pi+\sign\left[\myS{X,Y}(\om)\right] \nonumber\\ &\times&\arccos\left\{\frac{\myS{X}(\om)-\myS{Y}(\om)}{\sqrt{\bigl[\myS{X}(\om)-\myS{Y}(\om)\bigr]^{2}+4\left[\myS{X,Y}(\om)\right]^{2}}}\right\}.\qquad \label{optphase}
\end{eqnarray}
If in the experiment the phase of local oscillator has residual random fluctuations, the homodyne noise spectrum must be averaged over the distribution of the fluctuating phase $\varphi$, which we take as a Gaussian with variance $\Delta \varphi$, i.e.,
\begin{equation}
 \mySbar{d,\,\Delta\varphi}^{\varphi}(\om)=\frac{1}{\sqrt{2\pi}\bigl(\Delta\varphi \bigr)}\int_{-\infty}^{\infty}\mathrm{d}\varphi'\,
 \exp\Biggl\{-\frac{\bigl[\varphi'-\varphi \bigr]^{2}}{2\bigl(\Delta\varphi \bigr)^{2}}\Biggr\}\myS{\varphi '}(\om),
\end{equation}
which gives
\begin{eqnarray}\label{eq:avespe}
 \mySbar{d,\,\Delta\varphi}^{\varphi}(\om)&=&\frac{\myS{X}(\om)+\myS{Y}(\om)}{2}+\frac{\myS{X}(\om)-\myS{Y}(\om)}{2}\,\mathrm{e}^{-2 (\Delta \varphi)^2}\cos\bigl(2\varphi\bigr) \NML{+}\myS{X,Y}(\om)\,\mathrm{e}^{-2 (\Delta \varphi)^2} \sin\bigl(2\varphi\bigr),
 \end{eqnarray}
and as a consequence the optimal squeezing spectrum becomes
\begin{eqnarray}
 2\mySbar{d,\,\Delta\varphi}^{\mathrm{opt}}(\om)&=&\myS{X}(\om)+\myS{Y}(\om)-\mathrm{e}^{-2(\Delta\varphi)^{2}} \NML{\times}\sqrt{\bigl[\myS{X}(\om)-\myS{Y}(\om)\bigr]^{2}+4\bigl[\myS{X,Y}(\om)\bigr]^{2}}.\quad
\end{eqnarray}
Using the spectrum of the various noise sources, that is, shot noise spectrum $\myS{\mathrm{ in}}(\om) =1$, the thermal noise spectrum $\myS{\mathrm{ T}}(\om)=\left(\gamma_{\mathrm{m}}/\omega_{\mathrm{m}}\right)\omega\coth\Bigl(\hbar\omega/2\mathrm{k}_{\mathrm{B}}T\Bigr)$, the detuning noise spectrum $\myS{\dot{\phi}}(\om) $, and the amplitude noise spectrum $\myS{\epsilon}(\om)$, and Eq.~(\ref{eq:aout}), we get the general expressions of the output homodyne noise spectra of the amplitude quadrature $\myS{X}(\om)$, phase quadrature $\myS{Y}(\om)$, and of their correlation [$\myS{X,Y}(\om)$]. The explicit results are given in the Appendix.

\section{The experimental setup}

\begin{figure}[h]
 \centering
\includegraphics[width=0.475\textwidth]{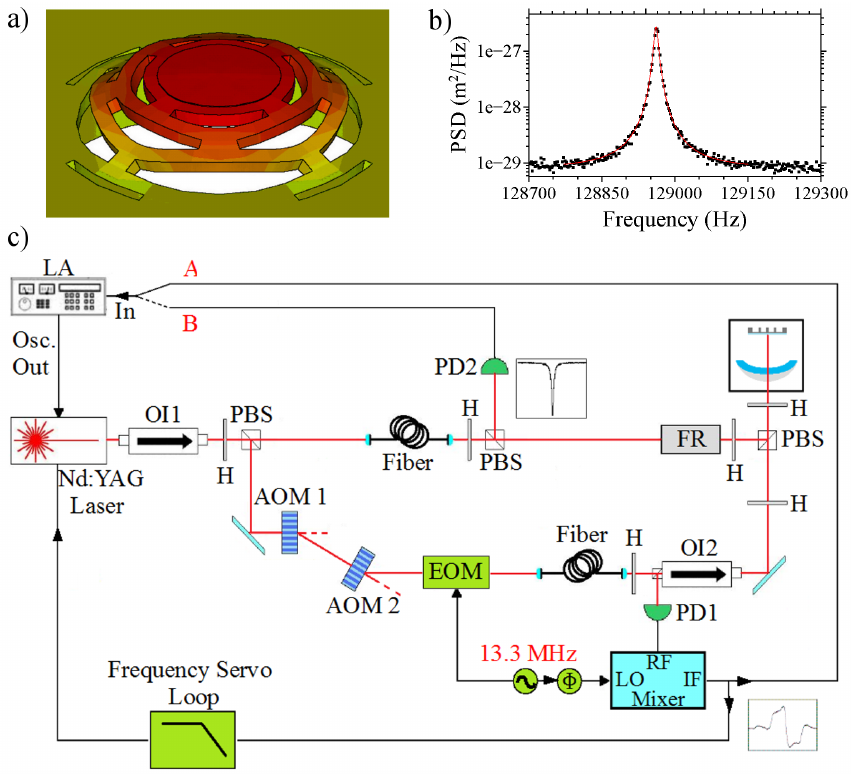}
\caption{(Color online) a) FEM image of the micromirror, showing the shape of the fundamental mechanical mode under study. The displacement is maximum in the red (dark gray) central mirror. b) Thermal displacement noise spectrum. c) Scheme of the experimental apparatus. Optical isolator (OI); acousto-optic modulator (AOM); electro-optic modulator (EOM); half-wave plate (H); polarizing beam splitter (PBS); Faraday rotator (FR); photodiode (PD); lock-in amplifier (LA); delay line for phase control ($\phi$). Black lines indicate the electronic part of the setup. We have omitted for clearness in the scheme an additional Michelson interferometer, monitoring the displacement of the micro-oscillator from its back side, and used for mechanical characterizations of the free oscillator (we have thus obtained, e.g., the spectrum b)).}
 \label{fig:appa}
\end{figure}

The mechanical oscillator used in the experiment belongs to a class of Micro-Opto-Mechanical-System specifically developed for having at the same time low optical and mechanical losses \cite{Serra2013}. The device has been realized in the Microfabrication Facility of FBK~\cite{fbk}. The reflective coating is obtained by the deposition alternate Ta$_2$O$_5$/SiO$_2$ quarter-wave layers for a total thickness of about 5.9 $\mu$m. Optical cavities with finesse up to 65\,000 have been realized with these devices. A structure made of alternate torsional and flexural springs supports the central mirror and allows its vertical displacement with a minimal internal deformation (see Fig.~\ref{fig:appa}a). This feature reduces the mechanical loss in the optical coatings and allows to reach quality factors up to 10$^5$. In this specific case the measurements were performed at room temperature, where the thermoelastic loss in the flexural springs prevents from achieving quality factors much
  surpassing $\sim10^4$ \cite{Serra2012a}.

The micro-oscillator is used as end mirror of a $0.57$~mm long Fabry-Perot cavity with a $50$~mm radius silica input mirror (transmissivity $\sim 50$~ppm) operating in a vacuum chamber at $10^{-3}$~Pa. The cavity finesse is $\mathscr{F}$~=~57\,000, (half-linewidth $\kappa/2\pi = 2.3$~MHz, with $\kappa_1 \simeq \kappa_2 \simeq \kappa/2$). The mechanical characteristics of the device are derived from the thermal noise spectrum measured with a polarization Michelson interferometer phase locked on a dark fringe (see Fig.~\ref{fig:appa}b) \cite{Serra2012}. The mechanical frequency is $\omega_m/2\pi =128\,961$~Hz, the mechanical quality factor $Q=16\,000$ and the effective mass $m = 1.35\times10^{-7}$~kg.

The experimental setup is sketched in Fig.~\ref{fig:appa}c. The light source is a cw Nd:YAG laser operating at $\lambda$=1\,064~nm. After a 40~dB optical isolator (OI1), the laser radiation is split into two beams. The first one (reference beam) is frequency shifted by means of two acousto-optic modulators (AOM) operating on opposite diffraction orders. A resonant electro-optic modulator (EOM) provides phase modulation at 13.3~MHz used for the Pound-Drever-Hall (PDH) detection scheme \cite{Pound1946,Drever1983}. The locking bandwidth is about 15 kHz and additional notch filters assure that the servo loop do not influence the system dynamics in the frequency region around the oscillator frequency. The second beam (pump), with a higher intensity, is employed to induce a strong optomechanical coupling as in usual schemes of pondermotive squeezing experiments.

Both beams are sent to the second part of the apparatus by means of single-mode, polarization maintaining optical fibers, overlapped with orthogonal polarizations in a polarizing beam-splitter and then mode-matched to the optical cavity with an efficiency above 95$\%$. Since the cavity is birefringent, the two beams are frequency-shifted with the AOM so that they both match the cavity resonance. The reflected reference beam, on its back path, is deviated by the input polarizer of a second optical isolator (OI2) and collected by a photodiode (PD1) for the PDH detection and laser frequency locking. The reflected pump beam, whose polarization is rotated by a double pass through a Faraday rotator (FR), is monitored by the photodiode~PD2.

In order to verify the frequency noise cancellation effect, we have studied the response function of the system to variations of the laser frequency in a narrow band around the mechanical resonance. The laser frequency is modulated by a sinusoid generated by the internal oscillator of a digital lock-in amplifier and applied to a piezoelectric transducer on the laser crystal. As we will discuss in the next Section, we have first characterized the frequency response of the system looking at the modulation in the PDH signal (signal A), when the pump beam is blocked. The detuning between the cavity resonance and the reference beam can be varied by adding an offset voltage to the PDH signal before the locking electronics. We have then repeated the measurement in the presence of the pump field, looking at the PDH signal and eventually at the reflected pump beam (signal B). In this configuration, the reference beam is resonant with the cavity and the detuning with respect to the pum
 p beam is varied by the acousto-optic modulators. The considered signals (A or B) are sent to the lock-in amplifier in order to extract the component (amplitude and phase) synchronous with the modulation. For a direct comparison between experimental results and theory, we have applied the following normalizations: (i) the frequency modulation depth (previously calibrated by a Michelson interferometer) is normalized to the half-linewidth of the cavity, measured independently from a frequency scan at very low laser power, giving $A_{\mathrm{in}}$; (ii) signal A is normalized to the peak-to-peak amplitude of the PDH, giving again (in the limit of small detuning and well resolved FM sidebands)  a displacement normalized to the cavity half-linewidth ($A_{\mathrm{PDH}}$); (iii) the modulated signal B is normalized to the height of the reflection dip, giving $A_{\mathrm{PDS}}$.

\section{Experimental demonstration of frequency noise cancellation}

We have experimentally verified the cancellation of frequency noise in a narrow band around the bare mechanical frequency $\omega_{\rm m}$ in two different ways: (i) looking at the dynamics of the PDH signal, which is approximately proportional to the cavity detuning and is therefore suitable to test the physics of the frequency noise cancellation described in Section \ref{IIIA}; (ii) measuring the intensity noise spectra of the field reflected by the cavity, which is the variable typically observed in ponderomotive squeezing experiments.

In both cases we have added a strong external frequency modulation in a region around $\omega_{\rm m}$ in order to better see the cancellation of the frequency fluctuations (otherwise overwhelmed by thermal noise). The calculated noise spectra can still be used to predict the experimental results, by just using a huge value of $\myS{\dot{\phi}}(\om) \simeq 10^{10}$ Hz (constant within the detection bandwidth) that becomes the dominant noise term.

\subsection{Detecting the fluctuations in the cavity detuning}

\begin{figure}[h]
 \centering
\includegraphics[width=0.475\textwidth]{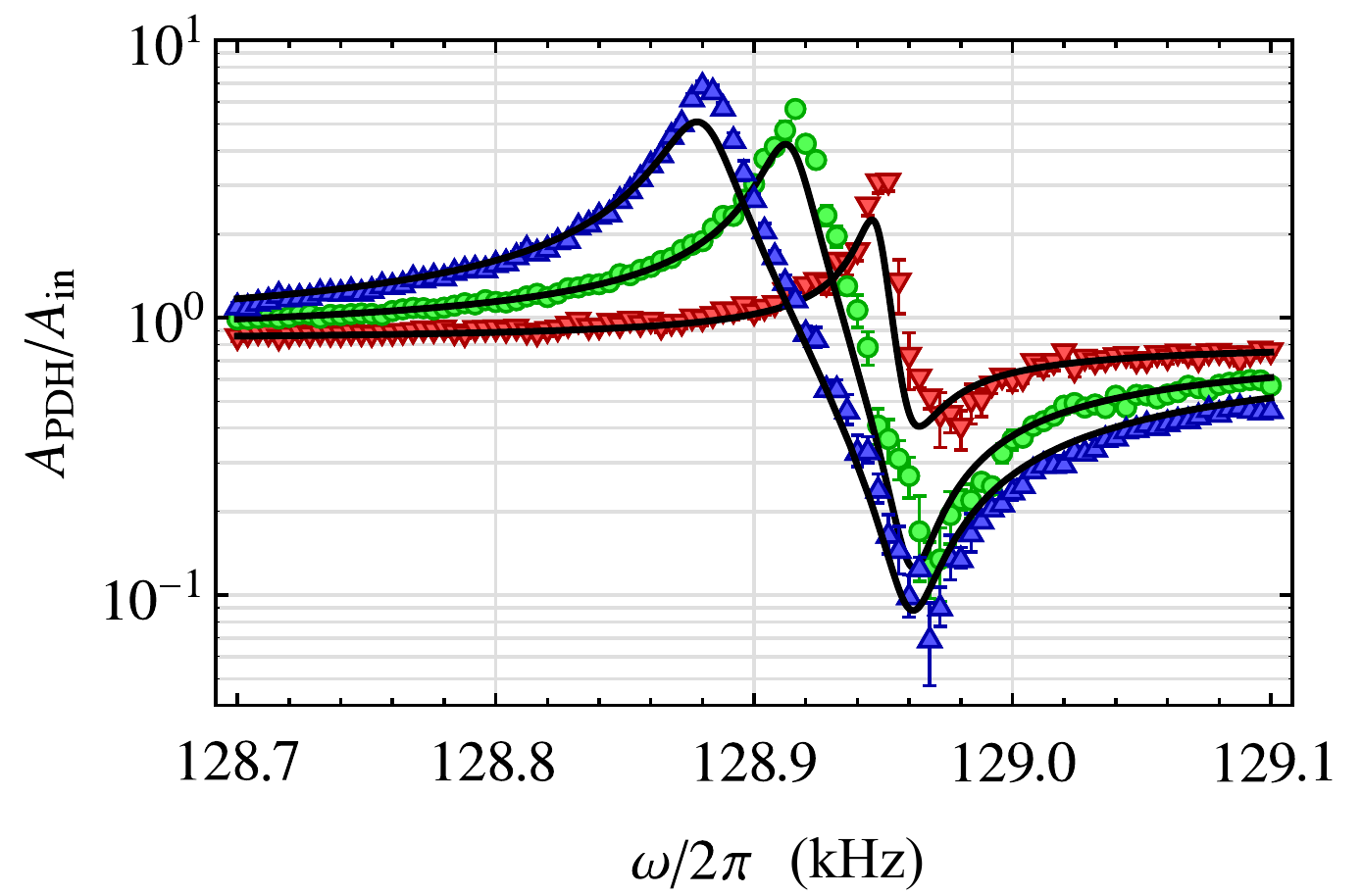}
\caption{(Color online) Normalized PDH response signal versus frequency with input power $\mathscr{P} = 0.09\,\mathrm{mW}$ and three different value of the detunings: $\varDelta = 0.0047 \kappa$ (red inverted triangles), $\varDelta =0.028 \kappa$ (green dots), $\varDelta =0.052 \kappa$ (blue triangles). Error bars express the statistical uncertainty on $\sim 5$ repeated measurements. The full lines correspond to the theory prediction with no fitting parameters, except for the addition of a supplementary detection noise due to the detection electronics (that determines the depth of the dips) and an overall adjustment of the vertical scale by $\sim20\%$ and fine adjustment of the driving power, both compatible with the uncertainty in the calibration of the experimental data.
 The other parameter values are those listed in Sec.~IV.}
 \label{fig:fase1}
\end{figure}

In the first case we have monitored the PDH signal, that can be considered as a measurement of the output phase quadrature~$Y^{\rm out}$. In Fig.~\ref{fig:fase1} we show the experimental data, normalized as explained in the previous Section, and the calculated spectrum normalized with respect to its value far from the mechanical resonance (corresponding to the signal at low frequency, zero detuning, and null optomechanical coupling). Experimental parameters are those given in Sec.~IV, with input power $\mathscr{P}=0.09\,\mathrm{mW}$.

\begin{figure}[h]
 \centering\includegraphics[width=0.475\textwidth]{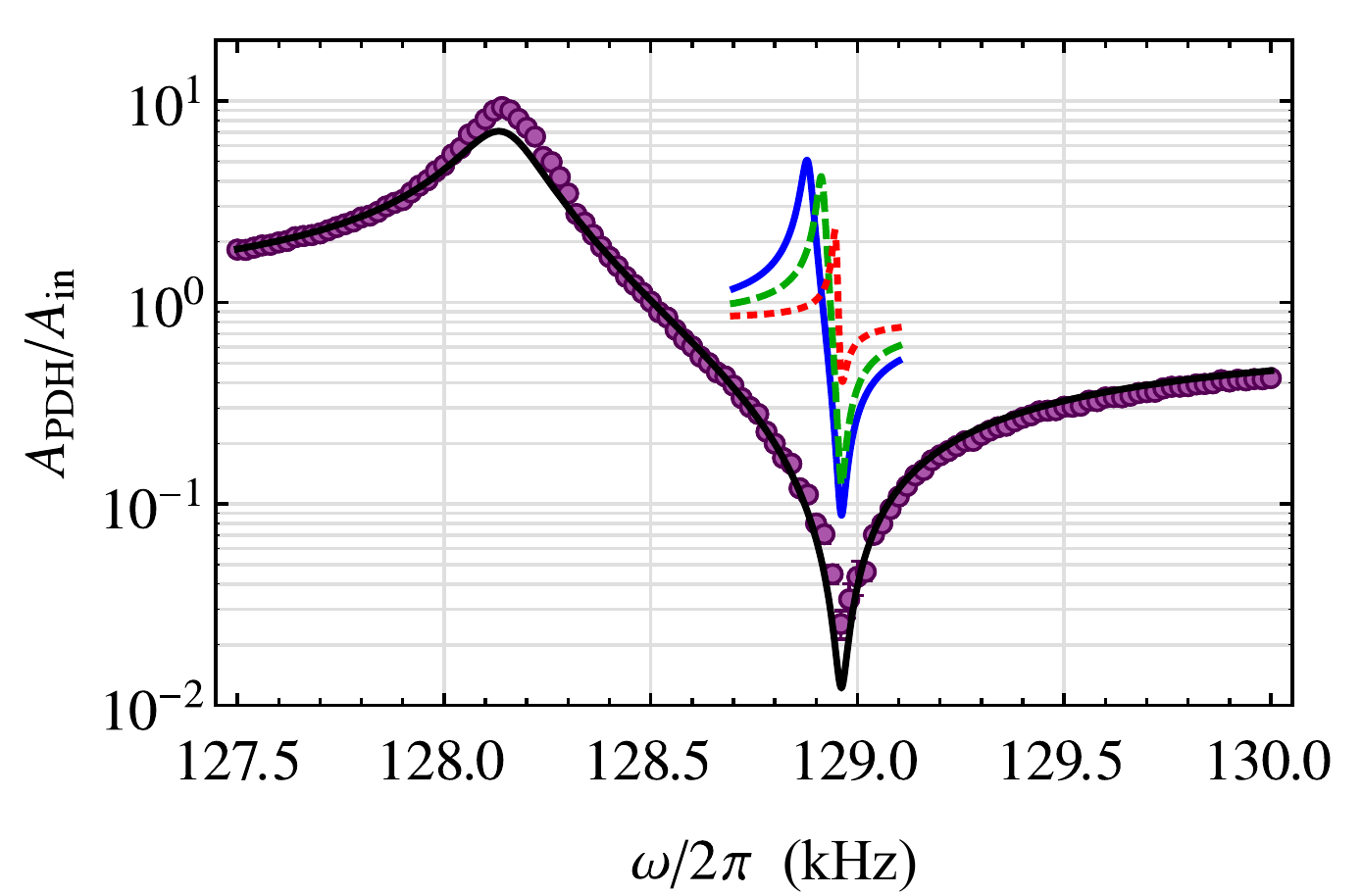}
\caption{(Color online) The same three theoretical curves of Fig.~\protect\ref{fig:fase1} for the normalized PDH signal versus frequency, together with additional data set (purple dots) corresponding to the PDH signal in the presence of an additional strong pump field which induces a much stronger modification of the effective susceptibility $\chi_\mathrm{ eff}(\omega)$, with input power $\mathscr{P}=1\,\mathrm{mW}$, and detuning $\varDelta =0.023 \kappa$. The full black line corresponds to the theory prediction with no fitting parameters, except for the addition of a supplementary detection noise due to the detection electronics.
The other parameter values are those listed in Sec.~IV.}
 \label{fig:fase2}
\end{figure}

Fig.~\ref{fig:fase1} shows that the dip is always exactly at $\omega = \omega_{\rm m}$ and that for increasing detuning the cancellation bandwidth increases and the dip is more pronounced, as it is also expected from the cancellation factor $\chi _{\rm eff}(\omega )/\chi _{0}(\omega )$ of Eq.~(\ref{eq:cancel}). In fact, the cancellation effect is larger when the effect of the optomechanical coupling on the modified effective susceptibility $\chi _{\rm eff}(\omega )$ is larger, i.e., for larger $\varDelta$ and $G$. This fact is confirmed by Fig.~\ref{fig:fase2} where, for comparison, we have added to the three curves of Fig.~\ref{fig:fase1}, the PDH signal associated with a weak probe resonant with the cavity in the presence of a much stronger and detuned pump field inducing a larger frequency shift associated with the optical spring effect. The additional purple dots in Fig.~\ref{fig:fase2} refer to this latter case and show a much deeper and visible cancellation effect, agai
 n centered at $\omega=\omega_m$. The associated full black curve is the theory prediction for the normalized PDH signal associated with the weak probe/reference field. More precisely, in this case, the probe signal is given by an equation analogous to Eq.~(\ref{eq:aout}), with coefficients $\nu_j(\omega)$ obtained by setting $\varDelta =0$ in Eqs.~(\ref{eq:nu}), but taking into account that both the optomechanical coupling $G$ and the effective susceptibility of Eq.~(\ref{chieffD}), are determined by the power and detuning $\varDelta$ associated with the strong pump field and not with the weak resonant probe. The agreement between theory and experiment is again very good in the case of a pump field with input power $\mathscr{P}=1\,\mathrm{mW}$, and detuning $\varDelta =0.023 \kappa$.

\subsection{Detecting the field reflected by the cavity}

\begin{figure}[h]
 \centering
\includegraphics[width=0.475\textwidth]{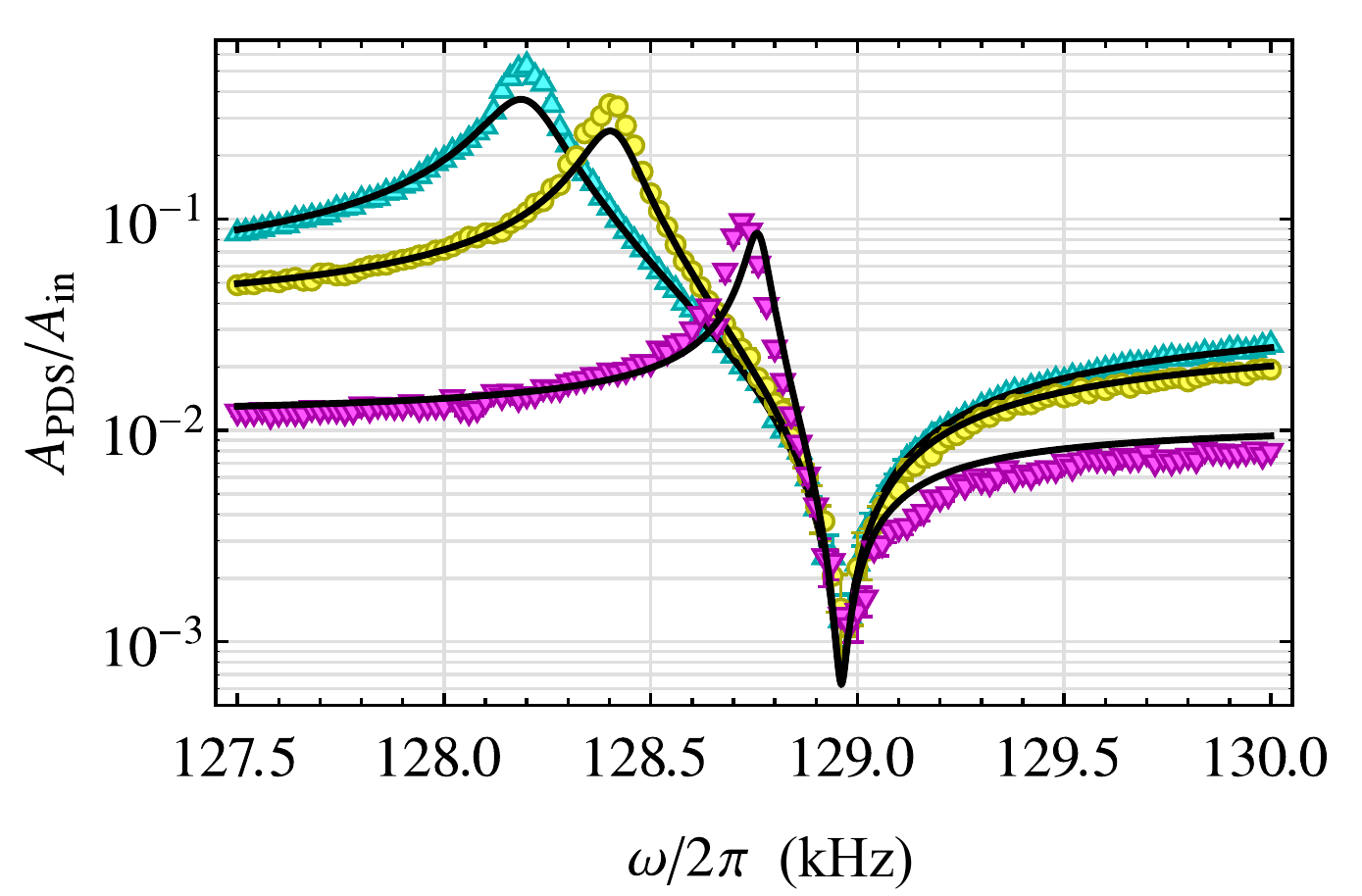}
\caption{(Color online) Normalized response function of the reflected field versus frequency for three different value of the detunings, $\varDelta \simeq 0.0056 \kappa $ (magenta inverted triangles), $\varDelta \simeq 0.015 \kappa$ (yellow dots), $\varDelta \simeq 0.021 \kappa$ (cyan triangles). The full lines correspond to the theory prediction, with no fitting parameters except for the addition of detection noise spectrum $\mySbar{\mathrm{dn}}= 10^7$. The other parameter values are those listed in Sec.~IV and with driving input power $\mathscr{P}\simeq 1\,\mathrm{mW}$.}
\label{fig:fase3}
\end{figure}

In the further stage of the experiment, we have detected the spectrum of the field reflected by the cavity, with average value
\begin{equation}\label{eq:reflected}
    E_\mathrm{R} = \sqrt{2\kappa_1}\alpha_s-\bar{a}_1^\mathrm{ in} = \frac{\EE_{0}}{\sqrt{2\kappa_1}}\frac{\kappa-2\kappa_1+\ii \De}{\ka+\ii\De},
\end{equation}
where we have used Eq.~(\ref{eq:ss_1th}) and $\bar{a}_1^\mathrm{ in}=\EE_{0}/\sqrt{2\kappa_1}$. This means in practice measuring the spectrum $\myS{d}^{\varphi_{\mathrm{R}}}(\om)$, where $\varphi_{\mathrm{R}} = \arctan\bigl[\De /\bigl(\kappa-2\kappa_1\bigr)\bigr]-\arctan\bigl(\De /\kappa\bigr)$ is the phase of $E_\mathrm{R}$. In Fig.~\ref{fig:fase3} we show the experimental data, normalized as explained in the previous Section, and the calculated spectrum. The theoretical data are normalized with respect to the depth of the reflection dip, such that the expression for the normalized detected signal can be written as
\begin{equation}
\frac{A_{\mathrm{PDS}}}{A_{\mathrm{in}}} = \frac{\kappa \lvert E_{\mathrm{R}}\rvert}{\sqrt{\myS{\dot{\phi}}}\left[\lvert \bar{a}_{1}^{\mathrm{in}} \rvert^{2} - \lvert  E_{\mathrm{R}}^{0} \rvert^{2}\right]}\sqrt{\myS{d}^{\varphi_{\mathrm{R}}}(\omega)+\mySbar{\mathrm{dn}}},
\end{equation}
where $\mySbar{\mathrm{dn}}$ is a constant detection noise due to the electronics.

This second experiment is more significative in view of the detection of ponderomotive squeezing because the reflected field is just the field where quadrature squeezing caused by radiation pressure becomes visible. Also in this noise spectrum we see the same features already underlined in the PDH signal: (i) the cancellation dip is exactly at the bare mechanical resonance $\omega_m$; (ii) the cancellation effect is more pronounced for larger couplings and detunings, i.e., for larger optical spring effect. Also in this case the theory prediction (full lines) well reproduces the data with no fitting parameters except for the addition of detection noise  $\mySbar{\mathrm{nd}}= 10^7$.

\section{Effect of noise cancellation for generating ponderomotive squeezing at low frequencies}

In this last section we show that the experimental setup studied above, if slightly improved, can be employed for generating ponderomotive squeezing at frequencies around $100$ kHz, i.e., considerably lower than those of Refs.~\cite{Safavi-Naeini2013,Purdy2013}, and that the cancellation mechanism illustrated above is of fundamental importance for the detection of squeezing. This can be seen by considering the prediction for the output homodyne noise spectrum defined by Eq.~(\ref{eq:spefi}) at a fixed phase $\bar{\varphi}$, which we have chosen as the optimal phase of Eq.~(\ref{optphase}) evaluated at the bare mechanical frequency $\bar{\varphi}=\varphi_\mathrm{ opt}(\omega_m)$. We have considered a slightly improved version of the setup, that is, the same optical cavity (i.e., same length and finesse), the same resonator mass and frequency, and considered only an improved mechanical quality factor, $Q = 10^5$ \cite{Serra2012a}, liquid He temperatures, $T = 4$~K, and larger input power, $\mathscr{P}\simeq 30\,\mathrm{mW}$.
The thick silicon micromirror employed here is able to manage high power at low temperatures, thanks to the favorable geometric factor (thicker connectors) and the high thermal conductivity of silicon at cryogenic temperature. In fact, as reported in Ref.~\cite{Serra2012} (see Fig.~4), the overall temperature drop induced by 1 mW of dissipated power is about 0.1 K, but the temperature increment of the mechanical springs is not more than 50 mK. As calculated in Ref.~\cite{Serra2012}, a dissipated power of 1 mW corresponds to an input power of about 14 mW. The input power value assumed here $\mathscr{P}\simeq 30\,\mathrm{mW}$ would provide a temperature increase with respect to the cold finger of few tenth of K, and a dissipated power of a couple of mW, well within reach of a standard liquid He cryostat.

In the device employed here, frequency noise is dominated by background noise and we have observed $\myS{\mathrm{bg}}\sim 10^{-34}\div 10^{-33}$ m$^2$/Hz in the $\sim100$ kHz region with up to $25$ mW of input power, and similar figures are shown by other groups~\cite{Arcizet2008}. We have in fact independently verified that laser frequency noise gives a negligible contribution which amounts to 1 Hz$^2/$Hz. For our predictions we take conservatively the upper limit $10^{-33}$ m$^2$/Hz, that is equivalent to the frequency noise $\myS{\dot{\phi}}=(\rmd \omega_{\rmc}/\rmd x)^2 \myS{\mathrm{bg}}\simeq (2\pi)^2\cdot300$ Hz$^2$/Hz. For what concerns laser amplitude noise $\ep$, the present apparatus, including an additional external noise eater, shows an excess noise 3 dB above the shot noise for a 30 mW laser beam (the work described in Ref.~\cite{Conti2000} has been recently extended in the $\sim100$~kHz range for this purpose). The excess amplitude noise could be further decreased by 20 dB using a standard (20 cm long, 30000 Finesse) filter cavity~\cite{Conti2003} and at this level, it would provide a negligible contribution to the output spectrum. As a consequence we have neglected the laser amplitude noise contribution in the predictions below, with the exception of Fig.~\ref{fig:zoom}.

\begin{figure}[h]
 \centering
\includegraphics[width=0.475\textwidth]{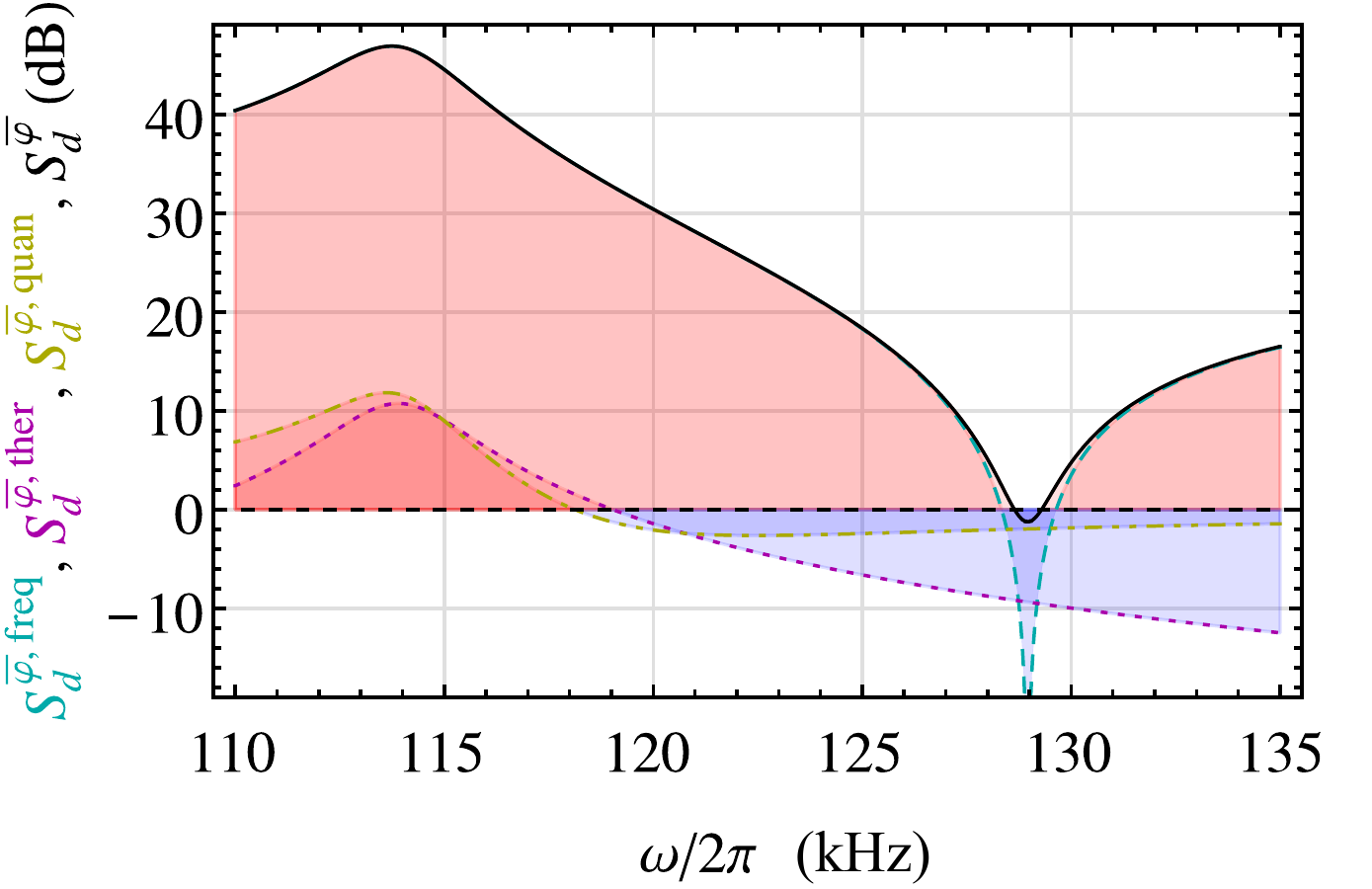}
\caption{(Color online) Homodyne noise spectrum $\myS{d}^{\bar{\varphi}}$ at the fixed phase $\bar{\varphi}=\varphi_\mathrm{ opt}(\omega_m)$ (black solid curve) together with its quantum noise contribution $\myS{d}^{\bar{\varphi},\,\,\mathrm{ quan}}$ (yellow dash-dotted curve), frequency noise contribution $\myS{d}^{\bar{\varphi},\,\mathrm{ freq}}$ (cyan dashed curve), and thermal noise contribution $\myS{d}^{\bar{\varphi},\,\mathrm{ ther}}$ (magenta dotted curve). Pondermotive squeezing is achieved in a narrow band around the bare mechanical frequency. Parameters are those of Sec.~IV, except that $Q = 10^5$, $T=4$ K, $\myS{\dot{\phi}}/(2\pi)^2=300$ Hz$^2$/Hz, $\myS{\ep}=0$, and with driving input power $\mathscr{P}= 30\,\mathrm{mW}$. The detuning is $\varDelta = 2\pi\cdot32\mathrm{kHz}$, corresponding to $0.014\kappa$.}
 \label{fig:pondsquee}
\end{figure}

\begin{figure}[h]
 \centering
\includegraphics[width=0.475\textwidth]{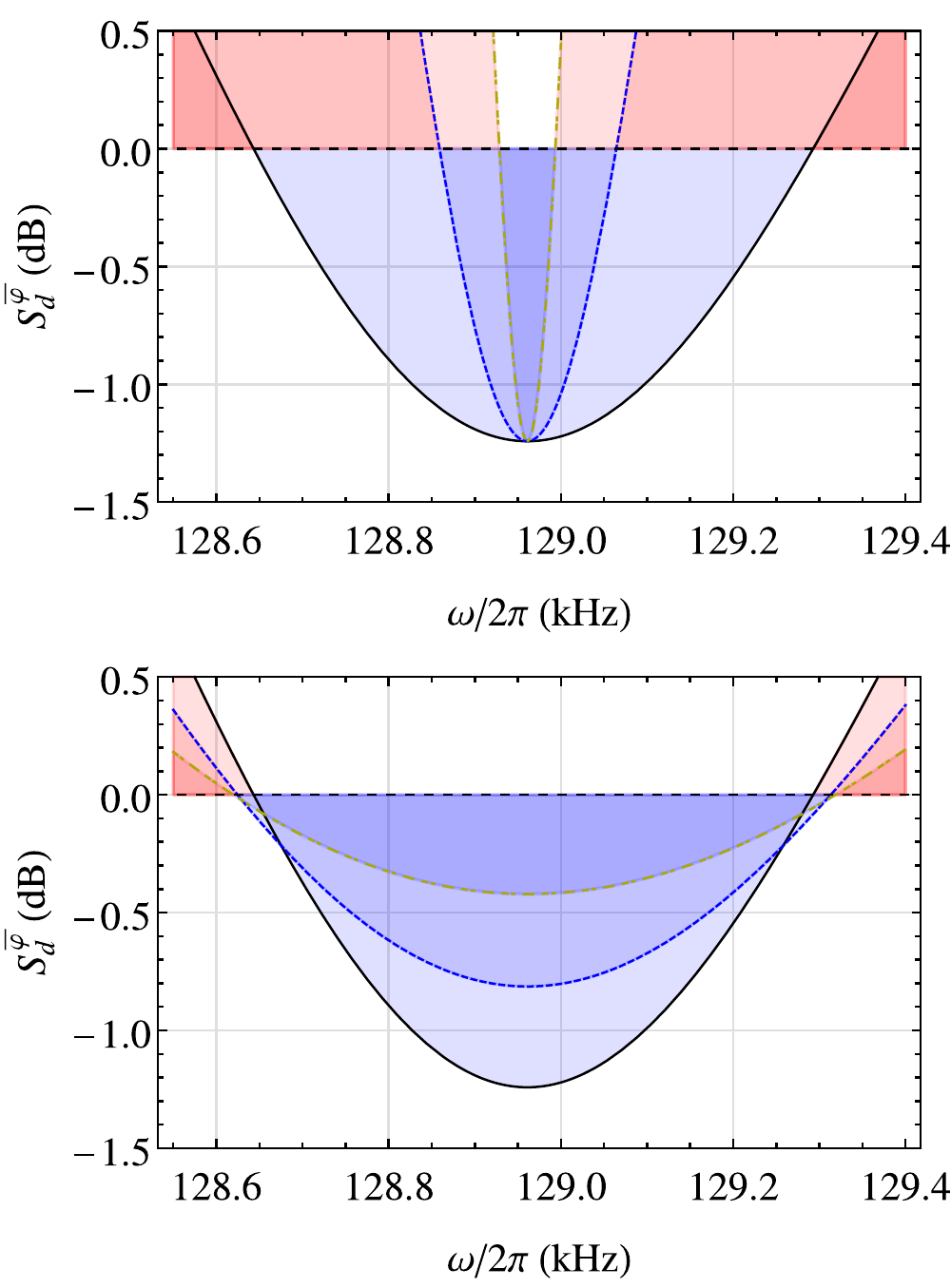}
\caption{(Color online) Enlarged view of the homodyne noise spectrum $\myS{d}^{\bar{\varphi}}$ around $\omega = \omega_m$. In the top figure $\myS{d}^{\bar{\varphi}}$ is shown at different values of frequency noise, $\myS{\dot{\phi}}/(2\pi)^2=3 \cdot 10^2$ Hz$^2$/Hz (black solid curve),
$\myS{\dot{\phi}}/(2\pi)^2=3 \cdot 10^3$ Hz$^2$/Hz (dotted blue curve), and $\myS{\dot{\phi}}/(2\pi)^2=3 \cdot 10^4$ Hz$^2$/Hz (yellow dot-dashed curve). In the bottom figure $\myS{d}^{\bar{\varphi}}$ is shown at different values of laser amplitude noise, $\myS{\ep}=0$ (black solid curve),
$\myS{\ep}= 0.2$ Hz/Hz (dotted blue curve), and $\myS{\ep}= 0.5$ Hz/Hz (yellow dot-dashed curve). This last value correspond to an excess noise 3 dB above the shot noise. The other parameters are those of Fig.~\protect\ref{fig:pondsquee}.}
 \label{fig:zoom}
\end{figure}

Fig.~\ref{fig:pondsquee} shows the homodyne noise spectrum $\myS{d}^{\bar{\varphi}}$ at phase $\bar{\varphi}= 178.6^{\circ}$, together with the various noise
contributions, i.e., the quantum noise, the frequency and thermal contributions. It is evident that one generates ponderomotive squeezing in a narrow bandwidth around the bare mechanical frequency $\omega_m$ only due to the frequency noise cancellation described above. In fact such noise is dominant everywhere except in this narrow band, where the detected homodyne spectrum is bounded below by the quantum noise contribution, in this set of parameters.

An enlarged view of the homodyne spectrum around $\omega_m$ is given by Fig.~\ref{fig:zoom}, where we show $\myS{d}^{\bar{\varphi}}$ at different values of the frequency noise $\myS{\dot{\phi}}$ (top) and of laser amplitude noise $\myS{\ep}$ (bottom). A larger $\myS{\dot{\phi}}$ implies narrowing the squeezing bandwidth, and we see that one can tolerate an appreciable amount of laser amplitude noise (see the figure caption for details). About 1 dB of squeezing is achievable in this parameter regime, comparable to that achieved in Refs.~\cite{Safavi-Naeini2013,Purdy2013}. Deeper and wider squeezing can be obtained for lower masses, higher $Q$, lower frequency noise and an over-coupled cavity (with $\kappa_2 \ll \kappa_1$) to reduce the effect of vacuum noise entering through optical losses.

\begin{figure}[h]
 \centering
\includegraphics[width=0.48\textwidth]{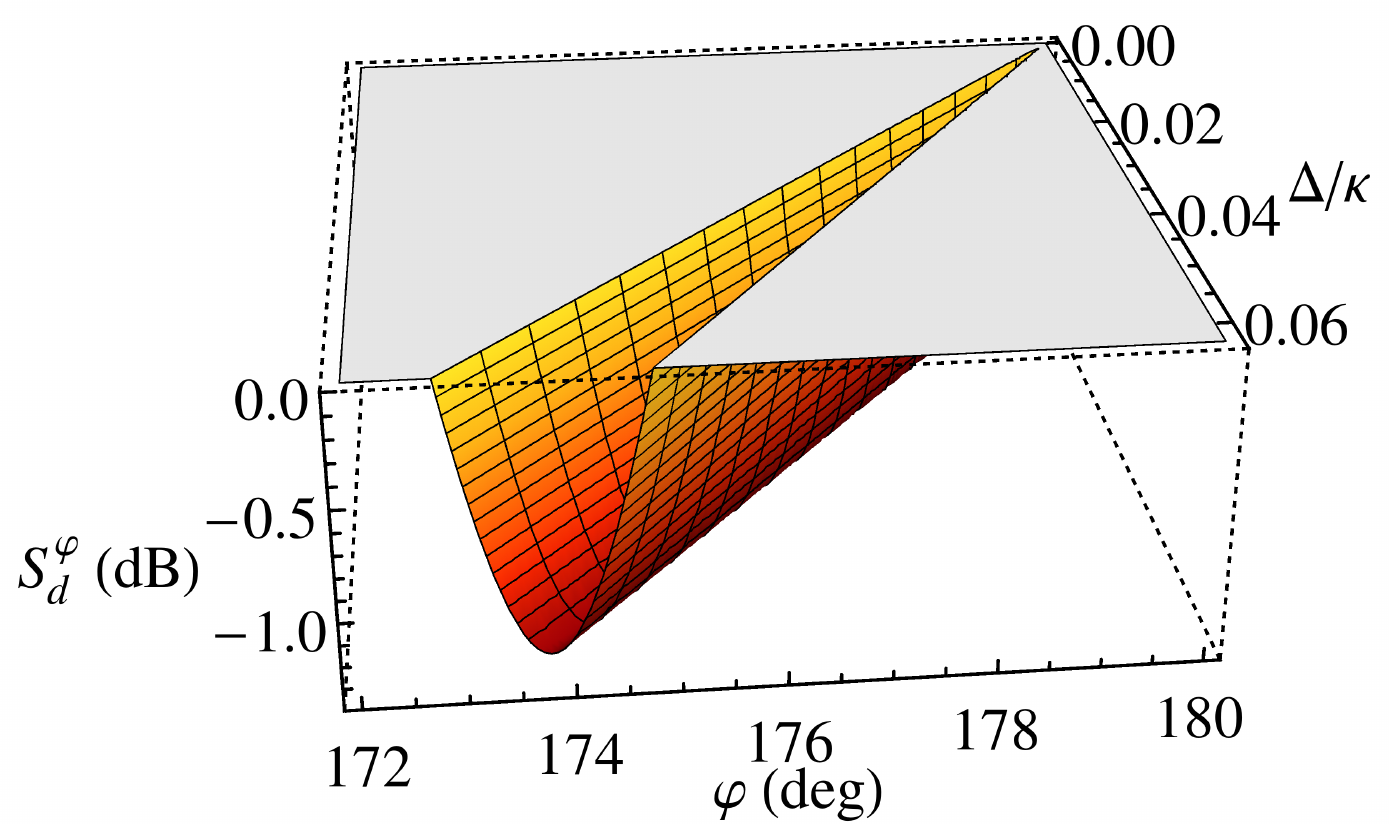}
\caption{(Color online) Homodyne noise spectrum $\myS{d}^{\varphi}$ at fixed frequency $\omega=\omega_{\mathrm{m}}$ as a function of the normalized detuning $\varDelta/\kappa$ and of the homodyne phase $\varphi$. Notation and parameters are those of Fig.~\protect\ref{fig:pondsquee}. The sub-shot noise region widens for increasing detunings and departing from $\varphi=0=\pi$.}
 \label{fig:detvarphi}
\end{figure}

\begin{figure}[h]
 \centering
\includegraphics[width=0.48\textwidth]{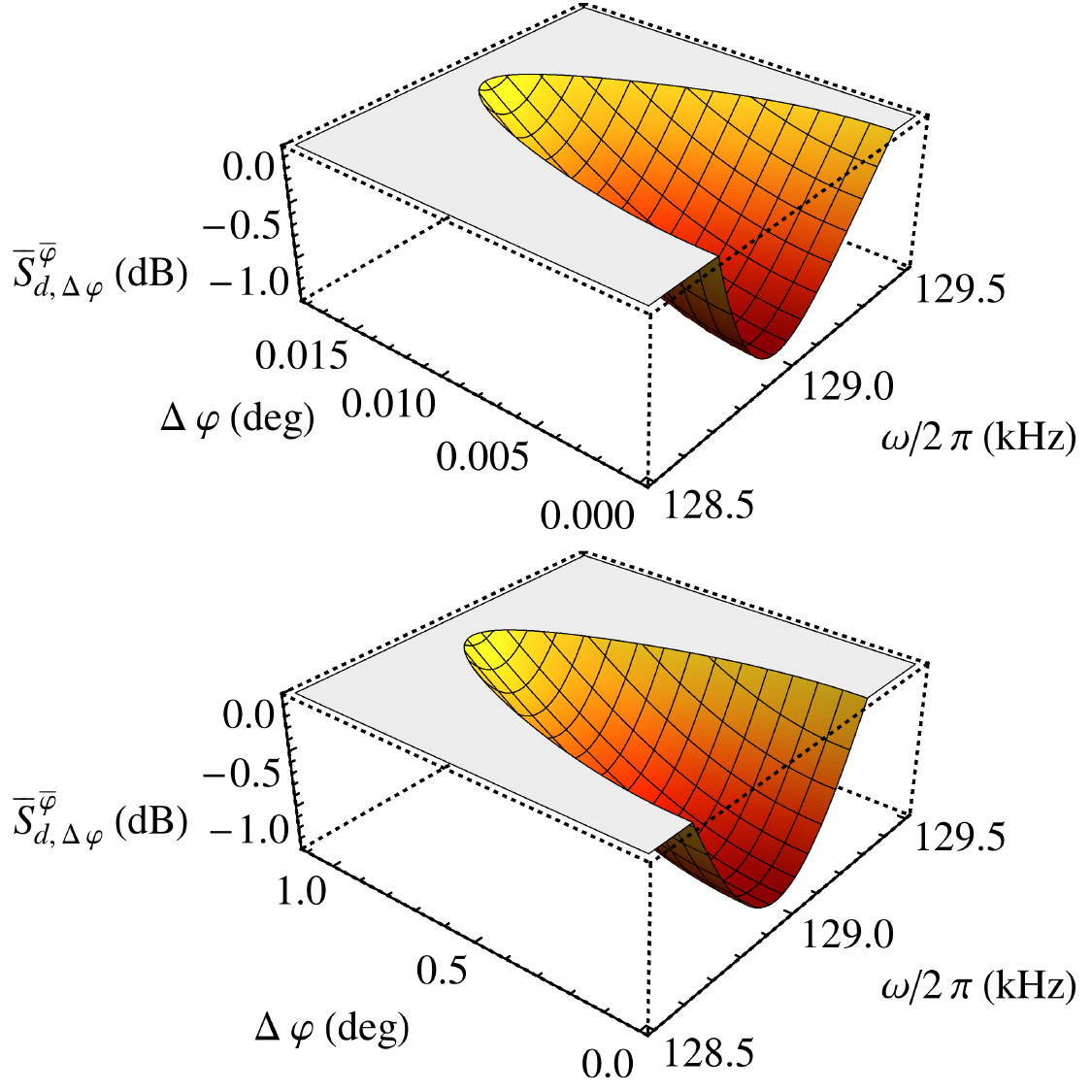}
\caption{(Color online) Averaged homodyne noise spectrum $\mySbar{d,\,\Delta\varphi}^{\bar{\varphi}}$ as a function of frequency $\omega/2\pi$ and detection phase uncertainty $\Delta\varphi$ at fixed detuning ($\varDelta/\kappa=10^{-3}$ in the upper panel and $\varDelta/\kappa=0.063$ in the lower panel) and fixed detection phase ($\bar{\varphi}=179.9^{\circ}$ in the upper panel and $\bar{\varphi}=173.8^{\circ}$ in the lower panel). Notice the different scales of the $\Delta\varphi$ axes. Notation and the other parameters are those of Fig.~\protect\ref{fig:pondsquee}.}
 \label{fig:avespe}
\end{figure}

The reason why frequency noise (in this case mostly due to background displacement noise) is so important is that it is transformed into strong intracavity laser intensity noise by the frequency-dependent resonance curve of the Fabry-P\'{e}rot. The conversion factor is roughly proportional to the derivative of the Lorentzian (at least in the \emph{bad cavity} limit), therefore typical calculations of achievable squeezing with realistic background noise are forced to consider very small detuning \cite{Marino2010,Serra2012}. A similar problem is found when aiming to measure quantum correlations induced by ponderomotive effect \cite{Verlot2011}. Such a small detuning means that the working point is quite close to the edge of the stability region, and that the requirement on the accuracy and stability of both the detuning and the homodyne phase are very tight. On the other hand, a further and crucial advantage of the frequency noise cancellation mechanism around $\omega=\omega_m$
  is that it allows to significantly relax the requirements on the stability and precision in the detuning $\varDelta$ and the detection phase $\varphi$ in order to get ponderomotive squeezing. This is illustrated in Fig.~\ref{fig:detvarphi}, where the homodyne noise spectrum $\myS{d}^{\varphi}$ at fixed frequency $\omega=\omega_{\mathrm{m}}$ is plotted as a function of the normalized detuning $\varDelta/\kappa$ and of the homodyne detection phase $\varphi$, with the same set of parameters of Fig.~\ref{fig:pondsquee}. The sub-shot noise region becomes wider and wider by increasing the detuning and, consequently, by departing from the phase of the amplitude quadrature $\varphi=0=\pi$. At larger detunings it is sufficient to stabilize the detection phase and the detuning itself at better than $1\%$ level in order to detect squeezing. On the contrary, closer to resonance $\varDelta =0$, the sub-shot noise region is much narrower and one has much more stringent stability requirem
 ents on $\varDelta$ and $\varphi$.
The fact that one can tolerate a significantly larger uncertainty in the detection phase $\varphi$ by operating around the noise cancellation point $\omega=\omega_m$ and at larger detunings can be seen also in the \emph{averaged} homodyne noise spectrum $\bar{S}_{d,\,\Delta\varphi}^{\varphi}(\om)$ of Eq.~(\ref{eq:avespe}) taking into account the presence of a detection phase uncertainty $\Delta\varphi$. This is shown in
Fig.~\ref{fig:avespe}, where $\bar{S}_{d,\,\Delta\varphi}^{\varphi}(\om)$ is plotted versus $\omega$ and $\Delta\varphi$ at fixed detuning and detection phase (namely, $\varDelta/\kappa=10^{-3}$ and the corresponding optimal phase $\bar{\varphi}=\varphi_\mathrm{ opt}(\omega_m)=179.9^{\circ}$ in the upper panel, and $\varDelta/\kappa=0.063$ and $\bar{\varphi}=\varphi_\mathrm{ opt}(\omega_m)=173.8^{\circ}$ in the lower panel). We see that at
small detunings squeezing vanishes already for an uncertainty $\Delta\varphi \simeq 0.015^{\circ}$, while at larger detunings ponderomotive squeezing is detectable up to a phase detection uncertainty $\Delta\varphi \simeq
1^{\circ}$. Further increase of the detuning is not convenient because at fixed input power, there is
an interval of values for $\Delta$ for which the system is unstable \cite{Genes2009}. Al larger values of the detuning the system is again stable but the achievable squeezing is lower. Similar results can be obtained by considering the uncertainty in the detuning $\varDelta$.

Achieving ponderomotive squeezing with the present optomechanical device presents some practical advantages with respect to the use of the setups of Refs.~\cite{Safavi-Naeini2013,Purdy2013}, which are characterized by higher mechanical frequencies and much lower masses. In this latter setups, radiation pressure effects are much stronger and therefore ponderomotive squeezing is easier to achieve. However, the mechanical frequency is much less stable and reproducible, because of significant stress drifts induced by thermal effects associated with optical absorption~\cite{Jockel2011}.
On the contrary, in the present thicker silicon micromechanical mirror, the expected temperature variation is just around 0.1 K in a cryogenic environment, due to the low thermal impedance of the device. This is confirmed by the fact that we could see no relevant drift in the mechanical resonance frequency by illuminating the sample from its back surface (where the light is partially absorbed by silicon) with laser power in the mW range.
Therefore the present optomechanical setup would be desirable in any application requiring squeezing in controlled and pre-determined frequency bands in the kHz range, such as for example, for improving the sensitivity of gravitational wave interferometers~\cite{LIGO2011}.

\section{Conclusions}

We have studied the generation and detection of ponderomotive squeezing in an optomechanical device formed by a Fabry-P\'{e}rot cavity with a micromechanical mirror~\cite{Serra2012} with mg mass. We have shown that ponderomotive squeezing is facilitated by a cancellation of the frequency noise around the bare mechanical resonance, which is caused by the destructive interference of the input frequency fluctuations and those induced by the mechanical response to the same frequency fluctuations. We have experimentally demonstrated such a cancellation mechanism, and we have also seen that, by slightly improving the apparatus, i.e., at lower temperatures, and higher mechanical quality factor, one could generate squeezing at kHz frequencies. Finally we have also shown that frequency noise cancellation allows to significantly relax the requirements on the stability and precision in the detuning and the detection phase in order to get ponderomotive squeezing.

\section{Acknowledgments}

This work has been supported by the European Commission (ITN-Marie Curie project cQOM), by MIUR (PRIN 2010-2011) and by INFN (HUMOR project).


\begin{appendix}

\section{General formulas for the homodyne noise spectra}

Using Eq.~(\ref{eq:aout}) one gets the following decomposition of homodyne noise spectra into quantum, frequency, amplitude, and thermal noise contributions
\begin{equation}\label{Xspectra}
    \myS{X}(\omega)=\myS{X}^\mathrm{ quan}(\omega)+\myS{X}^\mathrm{ freq}(\omega)+\myS{X}^\mathrm{ ampl}(\omega)+\myS{X}^\mathrm{ ther}(\omega),
\end{equation}
where
\begin{eqnarray}\label{Xdeco}
 \myS{X}^\mathrm{ quan}(\omega) &=& \tfrac{1}{2}\Bigl[\lvert\lambda_{12}^{+}(\omega)\rvert^{2} + \lvert\lambda_{12}^{+}(-\omega)\rvert^{2}\Bigr] \nonumber \\
 && + \Bigl[\lvert\lambda_{34}^{+}(\omega)\rvert^{2} + \lvert\lambda_{34}^{+}(-\omega)\rvert^{2}\Bigr] \kappa_{2}, \\
 \myS{X}^\mathrm{ freq}(\omega) &=& \biggl\{\bigl\lvert\lambda_{34}^{+}(\omega) + \lambda_{34}^{+,*}(-\omega)\bigr\rvert^{2}  \\
 && - 2\Re\Bigl[\Bigl(1+\mathrm{e}^{2\mathrm{i}\theta_{\varDelta}}\Bigr) \lambda_{34}^{+}(\omega) \lambda_{34}^{+}(-\omega)\Bigr]\biggr\}\lvert\alpha_{\mathrm
 s}\rvert^{2}\myS{\dot{\phi}}(\omega), \nonumber \\
 \myS{X}^\mathrm{ ampl}(\omega) &=& \bigl\lvert\lambda_{12}^{+}(\omega) + \lambda_{12}^{+,*}(-\omega)\bigr\rvert^{2}\myS{\epsilon}(\omega),  \\
 \myS{X}^\mathrm{ ther}(\omega) &=& \bigl\lvert\nu_{\mathrm{T}}(\omega) + \nu_{\mathrm{T}}^{*}(-\omega)\bigr\rvert^{2} \frac{\gamma_{\mathrm{m}}}{\omega_{\mathrm{m}}}\omega\coth\biggl(\frac{\hbar\omega}{2\mathrm{k}_{\mathrm{B}}T}\biggr),
 \end{eqnarray}
and
\begin{equation}
 \lambda_{ij}^{\pm}(\omega) \coloneqq \nu_{i}(\omega) \pm \nu_{j}^{*}(-\omega).
\end{equation}
In the same way,
\begin{equation}\label{Yspectra}
    \myS{Y}(\omega)=\myS{Y}^\mathrm{ quan}(\omega)+\myS{Y}^\mathrm{ freq}(\omega)+\myS{Y}^\mathrm{ ampl}(\omega)+\myS{Y}^\mathrm{ ther}(\omega),
\end{equation}
where
 \begin{eqnarray}\label{Ydeco}
 \myS{Y}^\mathrm{ quan}(\omega) &=& \tfrac{1}{2}\Bigl[\lvert\lambda_{12}^{-}(\omega)\rvert^{2} + \lvert\lambda_{12}^{-}(-\omega)\rvert^{2}\Bigr]  \nonumber \\
 && + \Bigl[\lvert\lambda_{34}^{-}(\omega)\rvert^{2} + \lvert\lambda_{34}^{-}(-\omega)\rvert^{2}\Bigr] \kappa_{2},  \\
 \myS{Y}^\mathrm{ freq}(\omega) &=& \biggl\{\bigl\lvert\lambda_{34}^{-}(\omega) - \lambda_{34}^{-,*}(-\omega)\bigr\rvert^{2}  \\
 &&  + 2\Re\Bigl[\Bigl(1+\mathrm{e}^{2\mathrm{i}\theta_{\varDelta}}\Bigr) \lambda_{34}^{-}(\omega) \lambda_{34}^{-}(-\omega)\Bigr]\biggr\}\lvert\alpha_{\mathrm
 s}\rvert^{2}\myS{\dot{\phi}}(\omega),\nonumber \\
 \myS{Y}^\mathrm{ ampl}(\omega) &=& \bigl\lvert\lambda_{12}^{-}(\omega) - \lambda_{12}^{-,*}(-\omega)\bigr\rvert^{2} \myS{\epsilon}(\omega), \\
 \myS{Y}^\mathrm{ ther}(\omega) &=& \bigl\lvert\nu_{\mathrm{T}}(\omega) - \nu_{\mathrm{T}}^{*}(-\omega)\bigr\rvert^{2} \frac{\gamma_{\mathrm{m}}}{\omega_{\mathrm{m}}}\omega\coth\biggl(\frac{\hbar\omega}{2\mathrm{k}_{\mathrm{B}}T}\biggr),
\end{eqnarray}
and
\begin{equation}\label{XYspectra}
    \myS{X,Y}(\omega)=\myS{X,Y}^\mathrm{ quan}(\omega)+\myS{X,Y}^\mathrm{ freq}(\omega)+\myS{X,Y}^\mathrm{ ampl}(\omega)+\myS{X,Y}^\mathrm{ ther}(\omega),
\end{equation}
where
\begin{eqnarray}\label{XYdeco}
\myS{X,Y}^\mathrm{ quan}(\omega) &=& \Im\bigl[\xi_{12}(\omega)\bigr] + 2\Im\bigl[\xi_{34}(\omega)\bigr] \kappa_{2} ,\\
\myS{X,Y}^\mathrm{ freq}(\omega) &=& 2\Im\Bigl[\zeta_{34}(\omega) - \Bigl(1+\mathrm{e}^{2\mathrm{i}\theta_{\varDelta}}\Bigr) \eta_{34}(\omega) \Bigr] \lvert\alpha_{\mathrm s}\rvert^{2}\myS{\dot{\phi}}(\omega),\qquad  \\
\myS{X,Y}^\mathrm{ ampl}(\omega) &=& 2\Im\bigl[\zeta_{12}(\omega)\bigr]\myS{\epsilon}(\omega), \\
\myS{X,Y}^\mathrm{ ther}(\omega) &=& 2\Im\bigl[\nu_{\mathrm{T}}(\omega) \nu_{\mathrm{T}}(-\omega)\bigr] \frac{\gamma_{\mathrm{m}}}{\omega_{\mathrm{m}}}\omega\coth\biggl(\frac{\hbar\omega}{2\mathrm{k}_{\mathrm{B}}T}\biggr),
\end{eqnarray}
and
\begin{eqnarray}
 \xi_{ij}(\omega) &\coloneqq& \nu_{i}(\omega)\nu_{j}(-\omega) + \nu_{i}(-\omega)\nu_{j}(\omega), \\
 \eta_{ij}(\omega) &\coloneqq& \nu_{i}(\omega)\nu_{i}(-\omega) - \nu_{j}^{*}(\omega)\nu_{j}^{*}(-\omega), \\
 \zeta_{34}(\omega) &\coloneqq& \bigl[\nu_{3}(\omega) + \nu_{4}(\omega)\bigr] \cdot \bigl[\nu_{3}(-\omega) + \nu_{4}(-\omega)\bigr].
\end{eqnarray}
Moreover $\theta_{\varDelta}=-\arctan\left(\varDelta/\kappa\right)$ is the argument of $\als$.

\end{appendix}

\bibliography{phase-noise-joint}

\end{document}